\documentclass[%
reprint,
amsmath,amssymb,
aps,
pra,
]{revtex4-2}

\usepackage{graphicx}% Include figure files
\usepackage{dcolumn}% Align table columns on decimal point
\usepackage{bm}% bold math
\begin{document}

%\preprint{APS/123-QED}

\title{
Topological Polarisation States
%Polarisation Torus, Hopf-link, and Topological Dirac Bosons
%Topology of Polarisation States
%Macroscopic Single-Qubit Operation for Coherent Photons
%Active Poincar\'e Rotator for Polarisation
%Dynamic $SU(2)$ Operation to Macroscopic Spin Angular Momentum of Coherent Photons
%Macroscopic $SU(2)$ Operation for Coherent Photons
%Dynamic $SU(2)$ Operation for Macroscopic Spin of Coherent Photons
%Active Poincar\'e Rotator for Polarisation
%Dynamic $SU(2)$ Operations to Macroscopic Spin of Photons in Coherent States
%Arbitrary Single Quantum Bit Operation for Macroscopic Coherent State of Photons
%Arbitrary Spin Rotation for Coherent Photons by Poincar\'e Rotator 
%Arbitrary $SU(2)$ Rotations of Macroscopic Quantum Bits in Coherent States
%Dynamic Control over Macroscopic Spin Angular Momentum of Coherent Photons
%Dynamic Control over Macroscopic Spin of Photons in Quantum Coherent States
%Dynamic Control of Macroscopic Quantum Bits Realised by Polarisation
%Arbitrary $SU(2)$ Rotations of Macroscopic Quantum Bits Spin Angular Momentum of Coherent Photons
%Quantum Operation , Quantum Coherent States
%Active Poincar\'e Rotator for Polarisation as Macroscopic Spin Angular Momentum of Coherent Photons
%Macroscopic Quantum States
%Single-Qubit Operation
}
%Active Poincar\'e Rotator for Polarisation as Macroscopic Spin of Coherent Photons}
% Force line breaks with \\

\author{Shinichi Saito}
 \email{shinichi.saito.qt@hitachi.com}
\affiliation{Center for Exploratory Research Laboratory, Research \& Development Group, Hitachi, Ltd. Tokyo 185-8601, Japan.}%Lines break automatically or can be forced with \\
%\author{Isao Tomita}%
%\affiliation{% Department of Electrical and Computer Engineering, National Institute of Technology, Gifu College, 2236-2 Kamimakuwa, Motosu, Gifu 501-0495, Japan.}%

\date{\today}% It is always \today, today, %  but any date may be explicitly specified

\begin{abstract}
Polarisation states are described by spin expectation values, known as Stokes parameters, whose trajectories in a rotationally symmetric system form a sphere named after Poincar\'e.
Here, we show that the trajectories of broken rotational symmetric systems can exhibit distinct topological structures in polarisation states.
We use a phase-shifter to form a polarisation circle (${\mathbb S}^1$), which interferes with the original input due to the phase change of the output state upon the rotation.
By rotating the circle using a rotator, the trajectories become a polarisation torus  (${\mathbb S}^1 \times {\mathbb S}^1$), which was experimentally confirmed in a simple set-up using passive optical components together with Mach-Zehnder interferometers.
We also discuss about realisations of other topological features, such as M\"obius strip, Hopf-links, and topological Dirac bosons with a bulk-edge correspondence.
\end{abstract}

%Max 600 characters in PRL, 500 words for PRA & PRB

\maketitle
%\tableofcontents

\section{Introduction}
Topology and quantum mechanics are inherently connected, and various exotic phenomena, such as quantum hall effect \cite{Ando74,Ando82,Klitzing80,Laughlin81,Kohmoto85,Hatsugai93}, spin hall effect \cite{Hirsh99,Murakami04,Wunderlich05}, topological insulator \cite{Haldane88,Kane05,Bernevig06,Konig07,Moore10}, and topological photonics \cite{Haldane08,Wang09,Hafezi13,Lu14,Price22}, were predicted theoretically \cite{Ando74,Ando82,Laughlin81,Kohmoto85,Hatsugai93,Hirsh99,Murakami04,Kane05,Bernevig06,Haldane08} and discovered experimentally \cite{Klitzing80,Wunderlich05,Konig07,Wang09,Hafezi13}.
These topological orders \cite{Berezinskii71,Kosterlitz73,Laughlin81,Thouless82,Haldane88,Kohmoto85,Hatsugai93,Wen04,Nagaosa99,Ezawa13} are different from thermodynamic spontaneous symmetry-breaking such as a superconducting phase-transition \cite{Nambu59,Anderson58,Goldstone62,Higgs64,Schrieffer71,Wen04,Nagaosa99}, which is characterised by opening an energy gap in the excitation spectrum to establish a long-range order \cite{Ginzburg50,Bardeen57,Schrieffer71}, while electronic or photonic states have continuous spectrum in a vacuum with full translational, time-reversal, and rotational symmetries \cite{Nambu59,Anderson58,Goldstone62,Higgs64,Schrieffer71,Coleman85,Shapere12,Wilczek12}.
In a topological material \cite{Ando74,Ando82,Laughlin81,Kohmoto85,Hatsugai93,Murakami04,Kane05,Bernevig06,Haldane08,Armitage18}, an energy gap is formed in the bulk as an insulator, which has different symmetry from that in a vacuum, such that the energy gap must be closed at the edge, whose state is topologically protected against structural imperfections as a highly conductive metal to accommodate massless Dirac fermions \cite{Laughlin81,Thouless82,Kohmoto85,Hatsugai93}.
This bulk-edge correspondence \cite{Laughlin81,Thouless82,Kohmoto85,Hatsugai93} is considered to be a generic feature of topological materials, whose topological invariants are Chern numbers \cite{Chern46}, obtained by integrating the Pancharatnam-Berry geometrical phase \cite{Pancharatnam56,Berry84,Tomita86,Cisowski22} of wavefunctions over the Brillouin zone \cite{Moore10,Hasan10,Qi11,Nakahara90}.
Thus, topological materials have unique topological band structures in the momentum space \cite{Moore10,Hasan10,Qi11}, rather than topological bonding configurations in the real space \cite{Shirakawa77,Kroto85,Iijima91,Tanda02,Novoselov04,Novoselov05,Fang09,Sunada12,Dabrowski-Tumanski17,Tomita19,Saito20}.

Here, we explore topological features in the polarisation space for spin states of coherent photons, that is, we consider topological aspects on polarisation states.
The polarisation state is described by an $SU(2)$ state, known as a Jones vector, which is a wavefunction of the spin state of photons \cite{Stokes51,Poincare92,Max99,Jackson99,Yariv97,Gil16,Goldstein11,Parker05,Chuang09,Hecht17,Pedrotti07,Grynberg10,Jones41,Fano54,Baym69,Sakurai67,Sakurai14,Saito20a,Saito20b,Saito20c,Saito20d,Saito20e,Saito21f,Saito22g,Saito22h}.
The wavefunction obviously has the amplitude and the phase, which are described by the polar angle ($\theta$) and the azimuthal angle ($\phi$) to show the average spin values as a vector to represent the state on the Poincar\'e sphere \cite{Stokes51,Poincare92,Max99,Jackson99,Yariv97,Gil16,Goldstein11,Parker05,Chuang09,Hecht17,Pedrotti07,Grynberg10,Jones41,Fano54,Baym69,Sakurai67,Sakurai14,Saito20a,Saito20b,Saito20c,Saito20d,Saito20e,Saito21f,Saito22g,Saito22h}.
In fact, we have recently demonstrated to realise an arbitrary polarised state by passive \cite{Saito22g} and active \cite{Saito22h} Poincar\'e rotators to execute an $SU(2)$ rotation of the Lie groups \cite{Stubhaug02,Fulton04,Hall03,Pfeifer03,Dirac30,Georgi99} in the combination of a $U(1)$ phase-shifter and a rotator.
As far as we are considering the coherent polarisation state in the power normalised configuration space, we can work on the Poincar\'e sphere with a unit radius ($r$) \cite{Yariv97,Gil16,Goldstein11}.
The Poincar\'e sphere with $r=1$ is equivalent to the Bloch sphere \cite{Baym69,Arecchi72,Narducci74,Sakurai14,Saito22g,Saito22h}, which means that trajectories of the polarisation states upon controlling the amplitude and the phase form a two-dimensional (2D) sphere (${\mathbb S}^2$, the surface of a ball in 3D space), which is topologically trivial with the genus ($g$) of zero with no hole, no knot, nor no link.

However, there is one noticeable difference between Poincar\'e and Bloch spheres, which is coming from Bose-Einstein and Fermi-Dirac statistics for photons and an electron, respectively \cite{Baym69,Sakurai67,Sakurai14,Wen04,Nagaosa99,Gil16,Goldstein11,Parker05,Fox06,Saito20a,Saito20d,Saito22h}. 
For photons, we can generate another photons with the same phase as that of an original photon upon stimulated emission process by using a polarisation independent Er-Doped Fibre Amplifier (EDFA) \cite{Yariv97,Gil16,Goldstein11,Parker05,Chuang09,Hecht17,Pedrotti07,Grynberg10,Saito20a,Saito20d,Saito22h}.
This corresponds to increase $r$ without changing the angles of $\theta$ and $\phi$, which is impossible to achieve for an electron due to the Pauli's exclusion principle \cite{Wootters82,Dieks82}.
The stimulated emission process requires a finite pumping power for the amplification, such that the process is not based on the norm-conserving unitary transformation \cite{Yariv97,Parker05,Saito22h}.
Therefore, the amplification of coherent photons is not violating with the no-cloning theorem \cite{Wootters82,Dieks82}, which prohibits copying of a quantum state by a unitary transformation, because the prerequisite of the notion for non-cloning is not satisfied by the injection of the pumping power.
Consequently, we consider a larger polarisation space, where the radius of the Poincar\'e sphere is not restricted to be unity but the polar coordinate of $(r,\theta,\phi)$ could span for the full 3D Euclidean space of Stokes parameters ${\bf S}=(S_1,S_2,S_3)$, which we propose to call as {\it the Stokes space}.
In the Stokes space, points with different $r$ can be distinguishable, even if $\theta$ and $\phi$ are the same.
This is not surprising, because we are dealing with signals with different intensities as for the means of digital communications \cite{Yariv97,Kikuchi16,Debnath18,Zhang21}, such as Quadrature-Amplitude-Modulation (QAM), Pulse-Amplitude-Modulation (PAM) for advanced multiplexing, and Dual-Polarisation Quadrature-Phase-Shift-Keying (DP-QPSK) \cite{Goi14,Doerr15}.
Stokes parameters \cite{Yariv97,Gil16,Goldstein11} can be described by energy per bit (pJ/bit) or power (mW).
Alternatively, they are also equivalent to the spin expectation values \cite{Saito20a,Saito20c}, which are obtained by the Dirac constant of $\hbar$ (the Plank constant of $h$, divided by $2\pi$), multiplied with the number of photons per second, passing through the area perpendicular to the direction of the propagation, with the spin pointing towards $x$, $y$, and $z$ directions, respectively.
In this paper, we use power for the dimension of Stokes parameters for simplicity.
In the Stokes space to take the difference in intensities into account, we can explore topologically non-trivial trajectories for the pulse streams generated from a device with broken rotational symmetries in polarisation states.

As an example of non-trivial polarisation state in the Stokes space, we first describe how to realise a polarisation torus, ${\mathbb T}^2  \cong {\mathbb S}^1 \times {\mathbb S}^1$, where $ {\mathbb S}^1$ is a 1D sphere to represent a polarisation circle, by using  passive optical components based on a simple representation theory of $U(2)  \cong U(1) \times SU(2)$ states to account for controlling the intensities by the Mach-Zehnder interferometers.
The comparison between the Stokes space and the Poincar\'e sphere is also discussed.
Then, we show our experimental results to confirm the theoretical expectations to realise the polarisation torus, which is realised as a non-trivial topological structure as trajectories in the Stokes space.
Novel non-transverse toroidal pulses have been recently observed out of meta-surfaces \cite{Zdagkas22}, while the mode of our polarisation torus is a standard fundamental mode in a single mode fibre and intensities together with phases are controlled to exhibit a torus as a set of points in the Stokes space.
We discuss about possibilities on realising more complex topological manifolds as polarisation states in the Stokes space such as M\"obius strip, Hopf-links, and topological Dirac bosons for the future.
We also discuss about the bulk-edge correspondence for these states, and show that the topological invariance for the proposed topological polarisation states is the Euler number and the genus in the Stokes space for spin expectation values, obtained by the Gauss-Bonnet theorem \cite{Nakahara90}, rather than the Chern number \cite{Chern46}, determined by the Pancharatnam-Berry phase \cite{Pancharatnam56,Berry84} in the $U(2)$ Hilbert space.

\section{Theoretical Designs}

We consider a propagation of light in a Single-Mode-Fibre (SMF) to make an argument specific, such that only the fundamental mode of a SMF is available for the propagation as a coherent wave \cite{Yariv97}. 
We have recently revisited the theoretical description \cite{Saito20a,Saito20d,Saito22g,Saito22h} for the coherent state of photons, emitted from a laser source, and confirmed that it should be treated as a many-body coherent state with the $SU(2)$ degrees of freedom for polarisation \cite{Jones41,Fano54,Baym69,Sakurai67,Sakurai14,Max99,Jackson99,Yariv97,Gil16,Goldstein11,Fox06}.
The coherent state \cite{Grynberg10,Fox06,Parker05,Saito20a,Saito22g,Saito22h} is characterised by the Gaussian distribution of the photon number, centred at the average number of photons per second, $\langle \hat{N} \rangle=N$, passing through the cross section of the SMF.
Practically, however, we do not have to employ creation and annihilation operators in coherent states for most of considerations in polarisation states \cite{Jones41,Max99,Jackson99,Yariv97,Gil16,Goldstein11}, and a wavefunction
\begin{eqnarray}
|N, \gamma, \delta \rangle
&=&
\sqrt{N} {\rm e}^{i \Phi}
| \gamma, \delta \rangle \\
&=&
\sqrt{N} {\rm e}^{i \Phi}
\left (
  \begin{array}{c}
    \cos (\alpha) \\
    {\rm e}^{i \delta}\sin (\alpha) 
  \end{array}
\right)
,
\end{eqnarray}
is enough to characterise the polarisation state on the Poincar\'e sphere \cite{Stokes51,Poincare92,Max99,Jackson99,Yariv97,Gil16,Goldstein11,Parker05,Chuang09,Hecht17,Pedrotti07,Grynberg10,Jones41,Fano54,Baym69,Sakurai67,Sakurai14,Saito20a,Saito20b,Saito20c,Saito20d,Saito20e,Saito21f,Saito22g,Saito22h}
, where $\Phi$ is the $U(1)$ phase of the orbital wavefunction, $\gamma=2\alpha$ is the polar angle measured from $S_1$, $\delta$ is the phase-shift measured from $S_2$, and $\alpha$ is the auxiliary angle for complex electric fields.
Here, we have used horizontal (H) and vertical (V) bases as for the fundamental states to describe the polarisation, and the normalisation of the wavefunction ($N$) is related to the power intensity of the ray as $P=\hbar \omega N=S_0$, where $\omega$ is the angular frequency and $S_0$ is the 0-th component of the Stokes parameter.
The $SU(2)$ nature of the polarisation is not affected by this normalisation, and it is straightforward to obtain the spin expectation values for photons as 
\begin{eqnarray}
\langle \hat{\bf S} \rangle
&=&
\hbar  \langle \hat{\bm \sigma} \rangle 
=
\hbar N
\left (
  \begin{array}{c}
    \cos \gamma \\
    \sin \gamma \cos \delta \\
    \sin \gamma \sin \delta 
  \end{array}
\right)
,
\end{eqnarray}
where the spinor vector of $\hat{\bm \sigma}=(\sigma_3,\sigma_1,\sigma_2)$ in HV bases is given by Pauli matrices
\begin{eqnarray}
\sigma_1=
\left(
  \begin{array}{cc}
0 & 1 \\
1 & 0
  \end{array}
\right),
\sigma_2=
\left(
  \begin{array}{cc}
0 & -i \\
i & 0
  \end{array}
\right) , 
\sigma_3=
\left(
  \begin{array}{cc}
1 & 0 \\
0 & -1
  \end{array}
\right), \nonumber \\
\end{eqnarray}
forming the Lie algebra of $\mathfrak{su}(2)$ \cite{Baym69,Sakurai14,Stubhaug02,Fulton04,Hall03,Pfeifer03,Georgi99,Saito20a,Saito22g,Saito22h}.
Thus,  the spin expectation values are related to the vectorial components of Stokes parameters as 
\begin{eqnarray}
{\bf S}
&=&
\omega \langle \hat{\bf S} \rangle
=
\hbar  \omega \langle \hat{\bm \sigma} \rangle
=
\hbar \omega N
\left (
  \begin{array}{c}
    \cos \gamma \\
    \sin \gamma \cos \delta \\
    \sin \gamma \sin \delta 
  \end{array}
\right)
,
\end{eqnarray}
in the unit of mW.
Alternatively, we can consider a normalisation based on the power as
\begin{eqnarray}
|P, \gamma, \delta \rangle
&=&
\sqrt{P} {\rm e}^{i \Phi}
| \gamma, \delta \rangle \\
&=&
\sqrt{P} {\rm e}^{i \Phi}
\left (
  \begin{array}{c}
    \cos (\alpha) \\
    {\rm e}^{i \delta}\sin (\alpha) 
  \end{array}
\right)
,
\end{eqnarray}
which we will employ, henceforth.

The exponential map \cite{Stubhaug02,Fulton04,Hall03,Pfeifer03,Dirac30,Georgi99} from the Lie algebra of $\mathfrak{su}(2)$ to the Lie group of $SU(2)$ is achieved by the unitary operator
\begin{eqnarray}
\hat{\mathcal{D}} ({\bf \hat{n}},\delta \phi) 
&=&
\exp 
\left (
-i 
\hat{\bm \sigma} \cdot {\bf \hat{n}}
\left (
\frac{\delta \phi}{2}
\right)
\right)\\
&=&
{\bf 1}
\cos
\left (
\frac{\delta \phi}{2}
\right)
-i 
\hat{\bm \sigma} \cdot {\bf \hat{n}}
\sin
\left (
\frac{\delta \phi}{2}
\right)
, 
\end{eqnarray}
where ${\bf \hat{n}}$ is the unit vector in the Stokes space and $\delta \phi$ is the rotation angle.
The operation  along the $S_3$ axis with ${\bf \hat{n}_3}=(0,0,1)$ is called as a rotator, and the operation along the $S_1$ axis with ${\bf \hat{n}_1}=(1,0,0)$ is called as a phase-shifter \cite{Yariv97,Gil16,Goldstein11,Saito20a,Saito21f,Saito22g,Saito22h}.
By combining a rotator and a phase-shifter, we could construct a Poincar\'e rotator, which allows an arbitrary rotation on the Poincar\'e sphere \cite{Saito21f,Saito22g,Saito22h}.
This corresponds to realise an $SU(2)$ rotation of the wavefunction of $|P, \gamma, \delta \rangle$, and the physical observable of $\langle \hat{\bf S} \rangle$ rotated in $SO(3)$.
The $SU(2)$ is an appropriate Lie group \cite{Stubhaug02,Fulton04,Hall03,Pfeifer03,Dirac30,Georgi99} for the polarisation\cite{Yariv97,Gil16,Goldstein11,Saito20a,Saito21f,Saito22g,Saito22h}, described by 2 complex numbers (${\mathbb C}$) in a normalised wavefunction \cite{Stokes51,Poincare92,Max99,Jackson99,Yariv97,Gil16,Goldstein11,Parker05,Chuang09,Hecht17,Pedrotti07,Grynberg10,Jones41,Fano54,Baym69,Sakurai67,Sakurai14,Saito20a,Saito20b,Saito20c,Saito20d,Saito20e,Saito21f,Saito22g,Saito22h}
, ensured by the determinant of unity, while $SO(3)$ is appropriate to describe a rotation of a vector, given by 3 real numbers (${\mathbb R}$) for spin expectation values.
The trajectories of the spin expectation values upon rotations form a sphere as a set of states, controlled by unitary operations \cite{Stokes51,Poincare92,Max99,Jackson99,Yariv97,Gil16,Goldstein11,Saito22g,Saito22h}.
The unitary operations means that we cannot change the energy of photons, such that the radius of the sphere is fixed, ideally.
In the reality for the practical implementation, we have finite insertion loss to control the polarisation states in the Poincar\'e rotators or conventional optical components, such as half/quarter wave-plates and rotators, which reduces the intensities.
Even in these cases, as far as the loss is not significantly dependent on the polarisation, the topology of polarisation states remained the same, such that we can analyse the polarisation state on the Poincar\'e sphere.
We could also introduce the gain by the polarisation independent EDFA \cite{Saito22h} to allow increasing the signal-to-noise ratio for polarimetry, but the polarisation states out of the devices are still accommodated on the sphere.
As far as we are dealing with the $SU(2)$ rotations on the sphere with or without polarisation independent loss or gain, the trajectories of the polarisation states are always on the sphere, which is topologically trivial.

Here, we consider to use another degree of freedom together with $SU(2)$ degrees of freedom, which is the $U(1)$ phase of $\Phi=kz-\omega t+\Phi_0$ for the orbital degree of freedom, where $k=2\pi/\lambda$ is the wavenumber for the wavelength of $\lambda$, and $z$ is the direction of the propagation along the SMF, and $\Phi_0$ is the initial phase.
The $U(1)$ phase plays no role, if we take the quantum average of spin states, as seen above, since it merely change the global phase of the wavefunction.
On the other hand, if we have another wave to compare the relative phase, the $U(1)$ phase could play a significant role.
For photons, this could be achieved simply by splitting the wave into 2 (or more) waves and introduce the relative phase change and re-combine to allow the interferences.
The $U(1)$ phase is related to the number of photons, such that the interference induces the changes in the number of photons, propagating to the SMF after the interference.
This process allows to control $P$ (or equivalentlly $N$), which corresponds to change the radius of the Poincar\'e sphere.
There will have several schemes to introduce the phase changes, and we consider one of the most simplest one, which just introduces the $SU(2)$ operation to one of the wave.
The SU(2) operation introduces the $U(1)$ phase change, which is observable upon interferences.
For example, one rotation on the Poincar\'e sphere of $SO(3)$ induces the sign change of the $SU(2)$ wavefuntion, because we expect
\begin{eqnarray}
\hat{\mathcal{D}} ({\bf \hat{n}},2\pi) 
&=&
-
{\bf 1}
,
\end{eqnarray}
which induces the destructive interference to the original input wave.
Mathematically, this is coming from the double-covering of $SU(2)$ to $SO(3)$, which is described as $SU(2)/{{\mathbb S}^0}  \cong SO(3)$, where ${\mathbb S}^0= \{ -{\bf 1}, {\bf 1} \}$ is the 0D sphere.
We need to rotate the amount of $4\pi$ to expect a complete rotation in $SU(2)$ with the identity of the operation, $\hat{\mathcal{D}} ({\bf \hat{n}},4\pi)={\bf 1}$.
Upon the phase change towards the interference, we can introduce both dynamic and adiabatic phases through hamiltonian (equivalently, rotators and phase-shifters) and geometrical configurations (Pancharatnam-Berry phase) \cite{Pancharatnam56,Berry84,Tomita86,Cisowski22,Moore10,Hasan10,Qi11,Nakahara90,Saito20a,Saito22g,Saito22h}.
Below, we will explain our practical deployment for realising non-trivial topological features as trajectories of polarisation states.

\begin{figure}[h]
\begin{center}
\includegraphics[width=8cm]{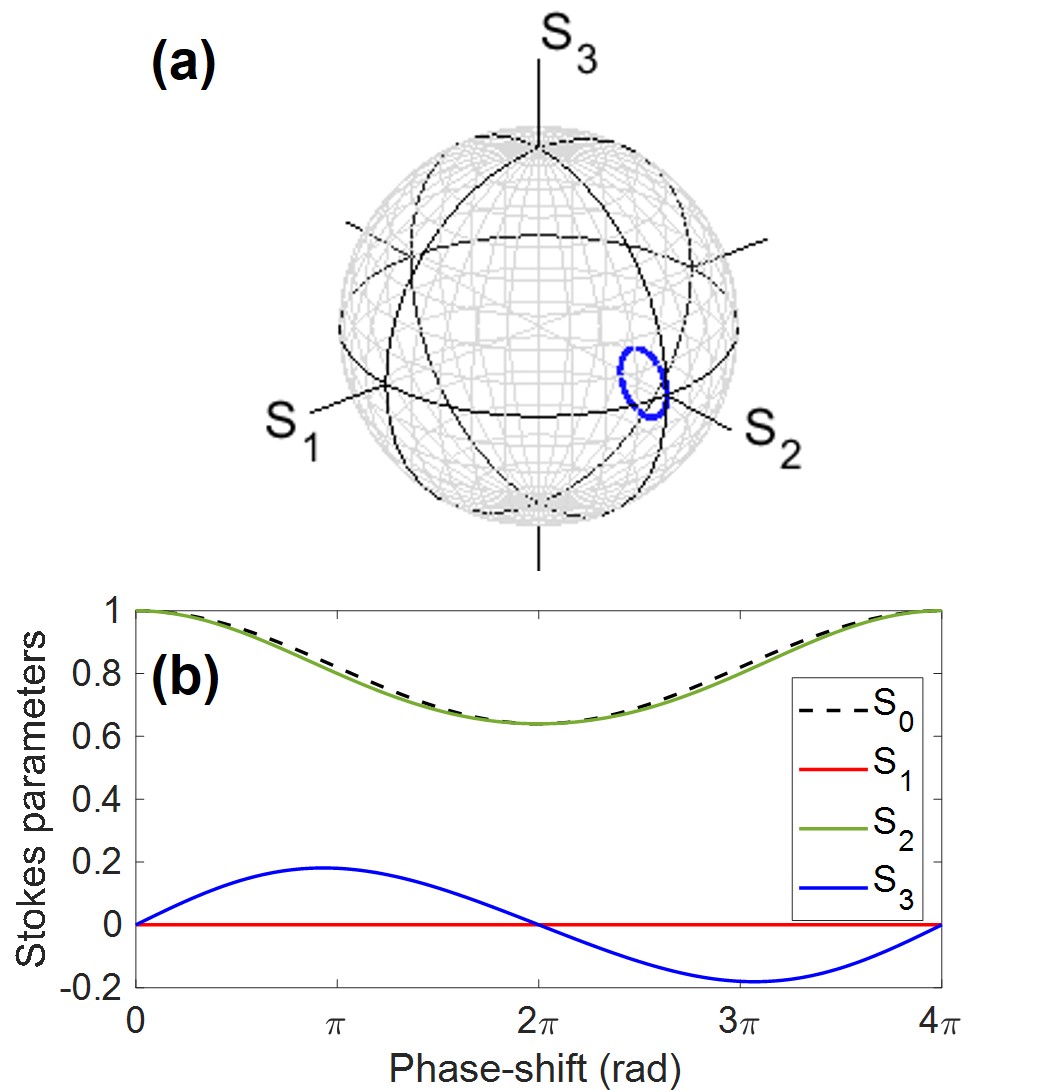}
\caption{
Calculated trajectories of polarisation states for the output from a polarisation interferometer.
The input state is prepared to be diagonally polarised with $S_2=1.0$ on the normalised Poincar\'e sphere.
(a) Stokes parameters, showing a polarisation circle (blue), which reduces its intensity upon the rotation.
(b) Stokes parameters against the phase-shift by an $SU(2)$ phase-shifter. 
The $2\pi$-rotation minimises the intensity in the diagonally polarised state and the $4\pi$-rotation is required to come back to the original input state.
}
\end{center}
\end{figure}

\subsection{Polarisation interferometer}

We explain the simple method to change the polarisation state together with the intensity in the Stokes space.
We assume specific experimental parameters to make the argument easy to understand, but it is straightforward to change parameters and to construct more generic theories.
First, we prepare the input wave with the power of $P_{\rm in}=1.5$ mW with the diagonally (D) polarised state, such that the input state of $|{\rm Input} \rangle $ is prepared as 
\begin{eqnarray}
\left | 
P_{\rm in}, \frac{\pi}{2}, 0 
\right \rangle
=
\sqrt{\frac{P_{\rm in}}{2}}
\left (
  \begin{array}{c}
1 \\
1  \end{array}
\right)
.
\end{eqnarray}
Then, we split the input wave into 2 waves by using a polarisation independent directional Fibre-to-Fibre Coupler (FFC).
We used the FFC of the splitting ratio of 90:10, which mean that the 90 \% of the signal is transmitted to the through port 3 and the 10 \% is coupled to the tap port 4, when we inserted from the input port 1, while the isolated port 2 is not used.
We define the coupling constant of $\alpha=0.1$ to account for the FFC, and the splitting is simply defined by a matrix operation, 
\begin{eqnarray}
\left (
  \begin{array}{c}
|{\rm Port \ 3} \rangle \\
|{\rm Port \ 4} \rangle 
  \end{array}
\right)
&=&
\left (
  \begin{array}{cc}
\sqrt{1-\alpha} & -\sqrt{\alpha} \\
\sqrt{\alpha} & \sqrt{1-\alpha} 
  \end{array}
\right)
\left (
  \begin{array}{c}
|{\rm Port \ 1} \rangle \\
|{\rm Port \ 2} \rangle 
  \end{array}
\right),
\nonumber \\
\end{eqnarray}
where 2 components of polarisation states $| P_{\rm in}, \pi/2, 0 \rangle$ is used for $|{\rm Port \ 1} \rangle$ and $|{\rm Port \ 2} \rangle$ is ${\bf 0}=(0,0)^{t}$, where $^{t}$ stands for the transpose of the vector.
The output from the tap port 4 is used to manipulate its polarisation state by a phase-shifter,
\begin{eqnarray}
\hat{\mathcal{D}} ({\bf \hat{n}_1},\delta \phi_{\rm p}) 
&=&
{\bf 1}
\cos
\left (
\frac{\delta \phi_{\rm p}}{2}
\right)
-i 
\hat{\sigma}_3
\sin
\left (
\frac{\delta \phi_{\rm p}}{2}
\right)
, 
\end{eqnarray}
which is the rotation along $S_1$ in the $S_2$-$S_3$ plane \cite{Saito22g}.
We defined the rotation angle of $\delta \phi_{\rm p}$, since this corresponds to the rotation along the poloidal direction, as we shall see below.
We have previously shown that we can construct a passive phase-shifter by a combination of 2 quarter-wave-plates (QWPs) and 2 half-wave-plates (HWPs) \cite{Saito22g}.
The amount of rotation in $SO(3)$ is determined by the physical angle of the rotation ($\delta {\it \Psi}_{\rm p}$) of one of the HWP as $\delta \phi_{\rm p}=4 \delta {\it \Psi}_{\rm p}$ \cite{Saito22g}.
Then, the output state of the tap port 4 becomes
\begin{eqnarray}
|{\rm Port \ 4^{\prime}} \rangle 
&=&
\hat{\mathcal{D}} ({\bf \hat{n}_1},\delta \phi_{\rm p}) 
|{\rm Port \ 4} \rangle 
,
\end{eqnarray}
while the output from the through port 3 is preserved to keep its polarisation state as $|{\rm Port \ 3^{\prime}} \rangle = |{\rm Port \ 3} \rangle$.
Then, we recombine the through port 3 and tap port 4 by the inverse arrangement, 
\begin{eqnarray}
\left (
  \begin{array}{c}
|{\rm Port \ 1^{\prime}} \rangle \\
|{\rm Port \ 2^{\prime}} \rangle 
  \end{array}
\right)
&=&
\left (
  \begin{array}{cc}
\sqrt{1-\alpha} & \sqrt{\alpha} \\
-\sqrt{\alpha} & \sqrt{1-\alpha} 
  \end{array}
\right)
\left (
  \begin{array}{c}
|{\rm Port \ 3^{\prime}} \rangle \\
|{\rm Port \ 4^{\prime}} \rangle 
  \end{array}
\right)
,
\nonumber \\
\end{eqnarray}
to expect that the port 1$^{\prime}$ is the main output, while the port 2$^{\prime}$ is not used.
Finally, the spin expectation values of $|{\rm Port}  \ 1^{\prime} \rangle$ are calculated, as shown in Fig. 1.

\begin{figure}[h]
\begin{center}
\includegraphics[width=8cm]{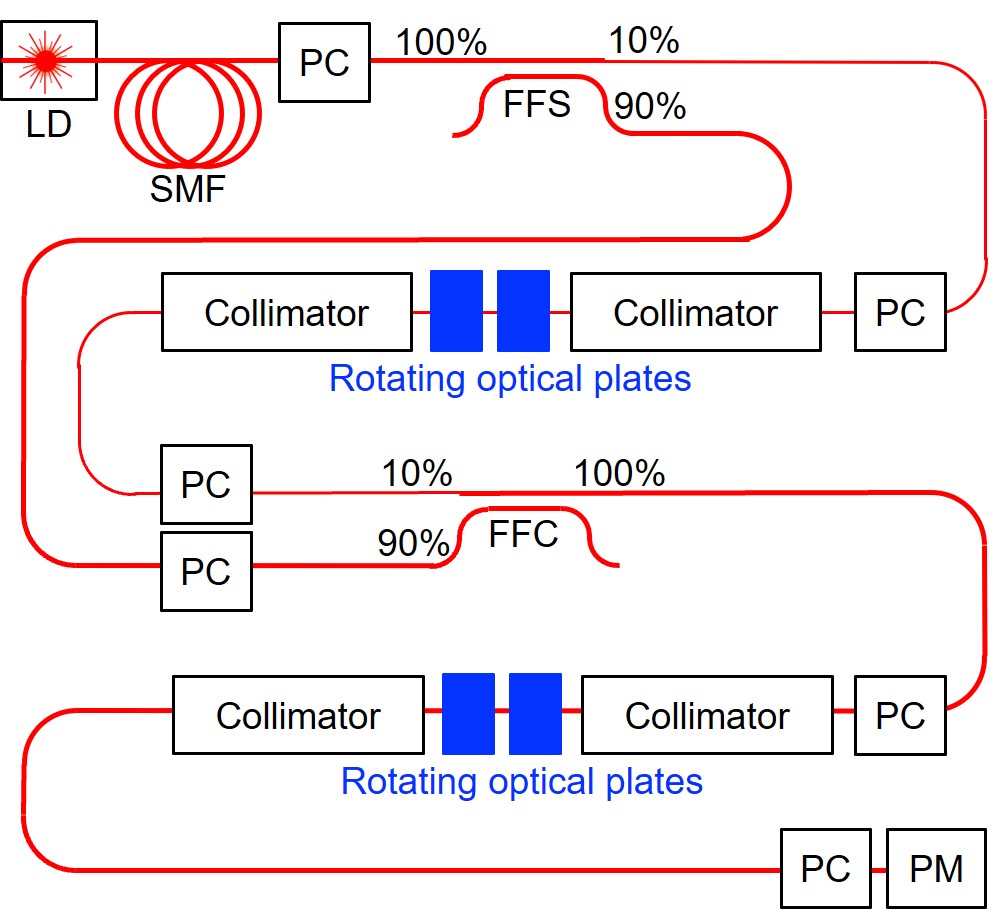}
\caption{
Polarisation interferometer to realise topologically non-trivial polarisation states in the Stokes space.
Rotating optical plates were used to change the spin states of $SU(2)$ as well as the orbital $U(1)$ phase for the bypassed wave.
These phase changes are observed upon interferences with the preserved input states, thus, allowing changes in intensities for the output state.
Abbreviations are as follows: 
LD: Laser Diode,  
SMF: Single Mode Fibre,  
PC: Polarisation controller,  
FFS: Fibre-to-Fibre Splitter
PM: Polarimeter, 
}
\end{center}
\end{figure}

We confirmed that the polarisation states are mostly located near the diagonally polarised state, $|{\rm D} \rangle$, since we assumed 90 \% of the input for the through port 3 is preserved.
Without the phase-shift of the tap port 4, we confirmed that the polarisation state and the intensity are not affected from the input state.
On the other hand, if we close look at the circular trajectory, we confirm that the trajectory is inside the original Poincar\'e sphere, which means that the radius, corresponding to the intensity, is successfully decreased.
The reduction in energy is confirmed in $S_0$ (Fig. 1(b)), which becomes the minium at $\delta \phi_{\rm p}=2\pi$ for the rotation of the tap port 4, where the the polarisation state is purely diagonally polarised.
This is coming from the double covering of $SU(2)$ to $SO(3)$, discussed above \cite{Saito20a}.
The circular rotation means that the input state of $|{\rm D} \rangle$ is coming back to the original state in the $SO(3)$ space, however, in the real physical space, it is enough for the complex electric field for the linearly-polarised diagonal state to rotate only for the rotation angle of $\pi$ to become the diagonal state upon the rotation.
This means that the electric field will be flipped to change the sign, which is not visible on the original Poincar\'e sphere.
On the other hand, this sign change is observable by using the portion of the original input wave, which is the role of the transmitted wave through port 3.
The destructive interference becomes maximum at $\delta \phi_{\rm p}=2\pi$, yielding the minimum of $S_0$.
The entire trajectory requires the $4\pi$-rotation to close, as expected for the $SU(2)$ group, and it became a polarisation circle ($\mathbb{S}^1$), which resides in the $S_2$-$S_3$ plane.
The operation process corresponds to a Mach-Zehnder interferometer with a polarisation control, associated with a phase-shift, and we name it as a {\it polarisation interferometer} (Fig. 2).
The poloidal polarisation circle is realised after the second FFC in Fig. 2.

We acquire a simple scheme to control the radius of the polarisation states, simply by mechanically rotating a HWP, which corresponds to change the intensity.
The total energy must be conserved upon the linear operations, and the reduced intensity is leaving from the isolated port 2, which is terminated.
Consequently, the output intensity could be reduced upon the insertion to our polarisation interferometer.
Thus, if we focus on the output waves, the system is not only based on unitary operations, but it allows a loss mechanism to reduce the radius, whose direction is perpendicular to the spherical directions, controlled by $\theta$ and $\phi$ for the polar-coordinate or by $\gamma$ and $\delta$ in the HV-bases.
In this sense, our system is non-Hermitian, and the original rotational symmetry of the polarisation state is also broken to expect a loss in particular direction.
In this way, we obtain a method to scan potentially entire a Euclidean coordinate in the Stokes space inside the original Poincar\'e sphere, spanned by the radius of the input wave.
Due to the 3D nature of the Stokes space, we can consider various topologically non-trivial structures, which we will explore below.

%For the 1 rotation over the Poincar\'e sphere, the phase change of $\delta \phi=2\pi$ is required, while the rotation of the polarisation ellipse in the real space is ${\it \Delta \Psi}=\pi$ \cite{Saito20a}.

% A push-pull configuration guarantees the SU(2) phase factor of -1 after the full rotation.
% Otherwise, in the 1-arm configuration, 1 rotation simply gives 1.
% We must be careful whether 1 rotation gives -1 or 1, even for the passive rotator.
% According to the SU(2) theory, the proper set-up should give -1 for the 1 rotation.
% Passive configuration will be fine, as is.
% For active control, it will be better to use a push-pull configuration.

\onecolumngrid

\begin{figure}[h]
\begin{center}
\includegraphics[width=16cm]{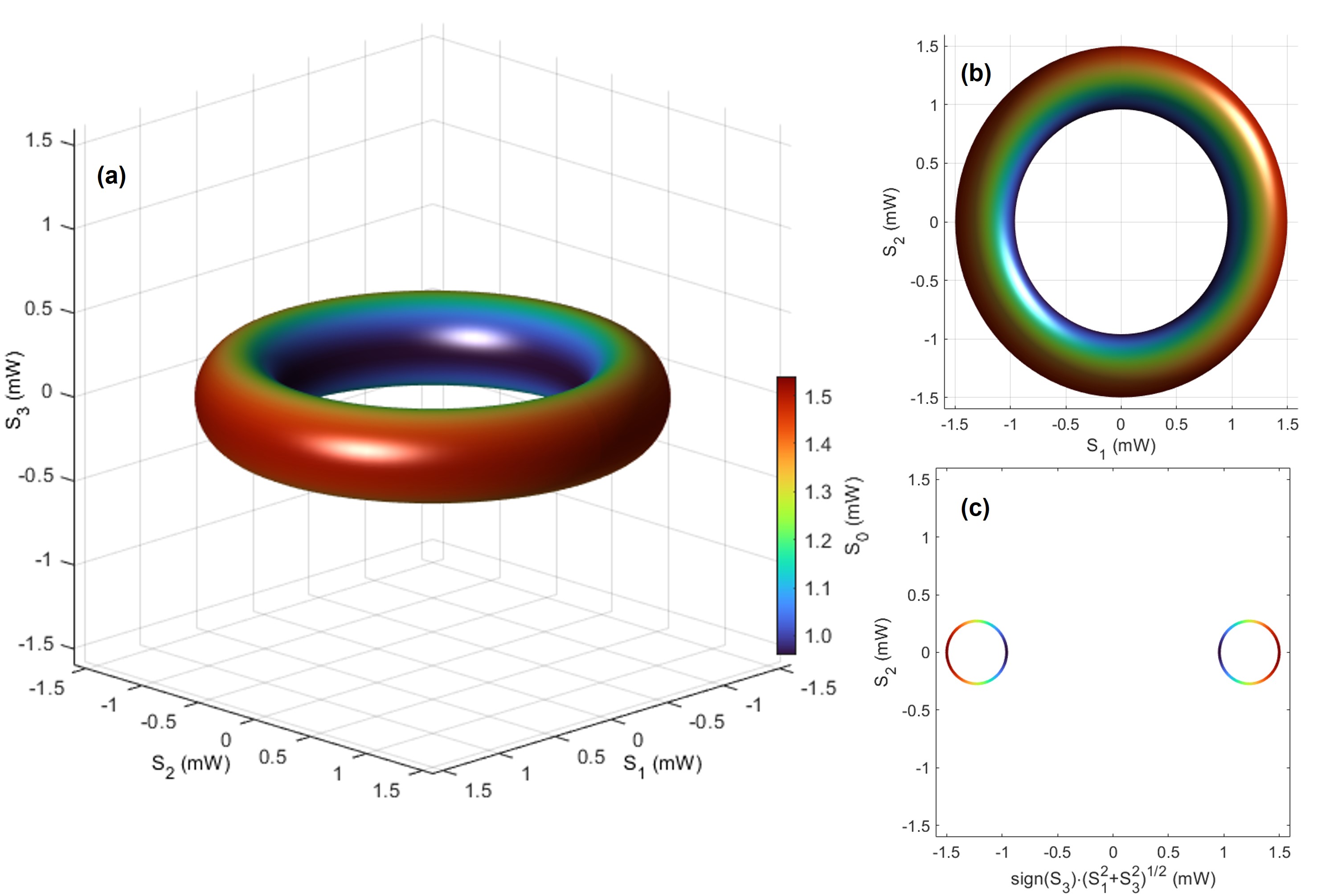}
\caption{
Polarisation torus in the Stokes space.
The input state to the polarisation interferometer is in the diagonally polarised state with the power of 1.5 mW.
The output power is controlled upon the interference, together with the polar angle and the phase, affected by rotating waveplates.
The Stokes parameters, $(S_0,S_1,S_2,S_3)$, of output states were calculated by using a simple representation theory of $U(2)  \cong U(1) \times SU(2)$.
(a) Stokes parameters are shown for various poloidal and toroidal rotation angles.
(b) Toroidal states, seen from the top of the $S_3$ axis.
(c) Poloidal states, as a cross section of the torus, perpendicular to the toroidal plane.
}
\end{center}
\end{figure}

\twocolumngrid

\subsection{Polarisation torus design}

As the first non-trivial topological structure of polarisation states, we explain how to construct a polarisation torus as a set of polarisation states in the Stokes space.
In topology, a torus is made of ${\mathbb T}^2  \cong {\mathbb S}^1 \times {\mathbb S}^1$, which requires topological groups to describe 2 circular rotations, orthogonal to each other.
In the previous subsection, we constructed rotators to describe the rotation along the poloidal direction, and therefore, we just need to add rotators to describe the rotations along the toroidal direction (Fig. 2).
This is achieved simply by applying a conventional rotator along the $S_3$ axis,
\begin{eqnarray}
\hat{\mathcal{D}} ({\bf \hat{n}_3},\delta \phi_{\rm t}) 
&=&
{\bf 1}
\cos
\left (
\frac{\delta \phi}{2}
\right)
-i 
\hat{\sigma}_2
\sin
\left (
\frac{\delta \phi_{\rm t}}{2}
\right)
, 
\end{eqnarray}
where $\delta \phi_{\rm t}$ is the amount of rotation along the toroidal direction, to the poloidal polarisation circle.
We have previously shown that two successive operations of HWPs are equivalent to a proper rotator operation to form the $SO(2)$ group \cite{Saito22g} rather than a pseudo rotator realised by one rotated HWP \cite{Yariv97,Gil16,Goldstein11}, and we just need to mechanically rotate one of the HWP to realise the target amount of rotation along the $S_3$ axis  (The last part of rotating optical plates in Fig. 2).
The output polarisation state after the operation becomes
\begin{eqnarray}
|{\rm Output} \rangle 
&=&
\hat{\mathcal{D}} ({\bf \hat{n}_3},\delta \phi_{\rm t}) 
|{\rm Port \ 1}^{\prime} \rangle 
,
\end{eqnarray}
and we can finally calculate the Stokes parameters from this output state (Fig. 3).

%\clearpage

\begin{figure}[h]
\begin{center}
\includegraphics[width=6cm]{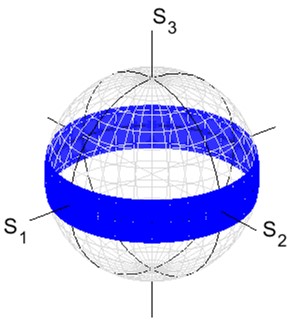}
\caption{
Polarisation torus collapsed on the Poincar\'e sphere.
The calculated Stokes parameters for the polarisation torus were mapped onto the Poincar\'e sphere after the normalisations at each point.
The torus becomes a belt with no information on the intensity.
}
\end{center}
\end{figure}

We confirmed that the calculated vectorial components of Stokes parameters form a polarisation torus (Fig. 3), and the intensities of the output satisfy $S_0=\sqrt{S_1^2+S_2^2+S_3^2}$, since we consider a coherent state.
As shown in the colour map of Fig. 3, $S_0$ depends on the location in the torus, and thus, we could realise a non-trivial topological structure with $g=1$.
For the definition of a torus, the states inside the torus are empty, which is confirmed in Fig. 3 (c).
It is also evident that the torus is compact and so closed as a set, such that even if we extend the amount of rotations for $\delta \phi_{\rm p}$ and $\delta \phi_{\rm t}$ beyond $4\pi$ and $2\pi$, respectively, we cannot generate new polarisation states outside the torus.
It is also evident from the construction of the torus, that $\hat{\mathcal{D}} ({\bf \hat{n}_1},\delta \phi_{\rm p})$ and $\hat{\mathcal{D}} ({\bf \hat{n}_3},\delta \phi_{\rm t})$ from $U(1) \cong  {\mathbb S}^1$ groups, such that the 2 successive operations could be performed by 1 operation and the inverse of a rotation could be defined.
This was also true for a spherically symmetric Poincar\'e sphere, where the rotator and the phase-shifter are physical realisation of Lie group operations of $SU(2)$ \cite{Saito20a,Saito22g,Saito22h}.
In the present case, the torus is realised upon the interference to reduce its intensity, which is non-reversal process, such that these rotational operations must be completed before the interference is taken place.
As far as rotations are made before the interference, we can consider alternative operations.
For example, we could first rotate the polarisation state of the input along the $S_3$ axis to control along the toroidal direction, and then, we could split into 2 waves to allow the poloidal rotation.

\begin{figure}[h]
\begin{center}
\includegraphics[width=8cm]{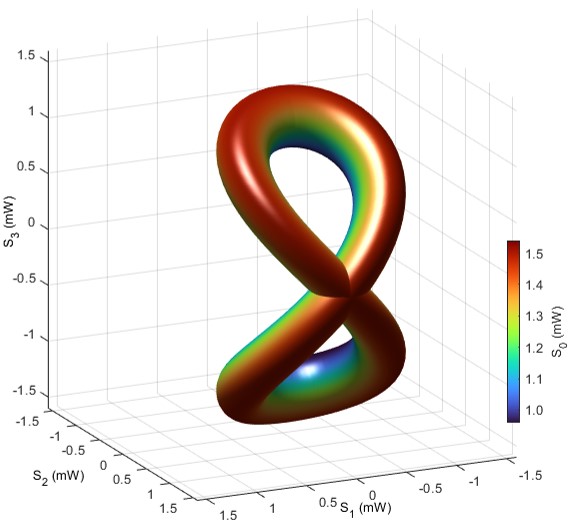}
\caption{
Coupled polarisation torus.
Stokes parameters were calculated, using a rotated quarter-wave-plate, applied to a polarisation circle.
2 holes are visible, but the structure is based on 2 connected toruses with genus of 1.
}
\end{center}
\end{figure}

%\clearpage

We have also plotted the calculated Stokes parameters on the Poincar\'e sphere after normalisation at each output state (Fig. 4).
In this case, we cannot discuss differences in relative intensities and the trajectories of the torus collapsed to be a belt on the Poincar\'e sphere with no width (Fig. 4).
The mapping corresponds to a projection from $\mathbb{R}^3$ to $\mathbb{S}^2$, and the information on radius, corresponding to $N$ or $P$ will be gone. 
One can always consider this mapping from a torus to a belt, if he/she wants, but this ends up considering non-trivial topological structures in $\mathbb{R}^3$.
We will come back to this point, when we discuss about the Chern number \cite{Chern46}, the Pancharatnam-Berry phase \cite{Pancharatnam56,Berry84}, and the Gauss-Bonnet theorem \cite{Nakahara90} towards the end of this paper.
Here, we emphasize the non-trivial topological feature has appeared in the Stokes space, which uses $\mathbb{R}^3$ in $SO(3)$ rather than $\mathbb{C}^2$ for $U(2)$ wavefunctions, and for bosons, it is meaningful to consider the difference in $N$ due to their Bose-Einstein statistics.

As shown in Fig. 5, we have also calculated polarisation torus by using the rotated QWP as the final rotation in Fig. 2, instead of the rotator.
In this case, the trajectories by a rotated QWP for the input of the D-state become '8'-like structure with 2 holes \cite{Saito22g}, such that the rotation of the polarisation circle of Fig. 1 (a) upon the rotated QWP form 2 torus (Fig. 5).
Unfortunately, this structure is not a tours of $g=2$, but it is simply 2 toruses of $g=1$ overlapped each other, since the simple moves of the circle intersects near the D-state.
Therefore, it is not topologically distinguishable to the torus of $g=1$, shown in Fig. 3.

\onecolumngrid

\begin{figure}[h]
\begin{center}
\includegraphics[width=16cm]{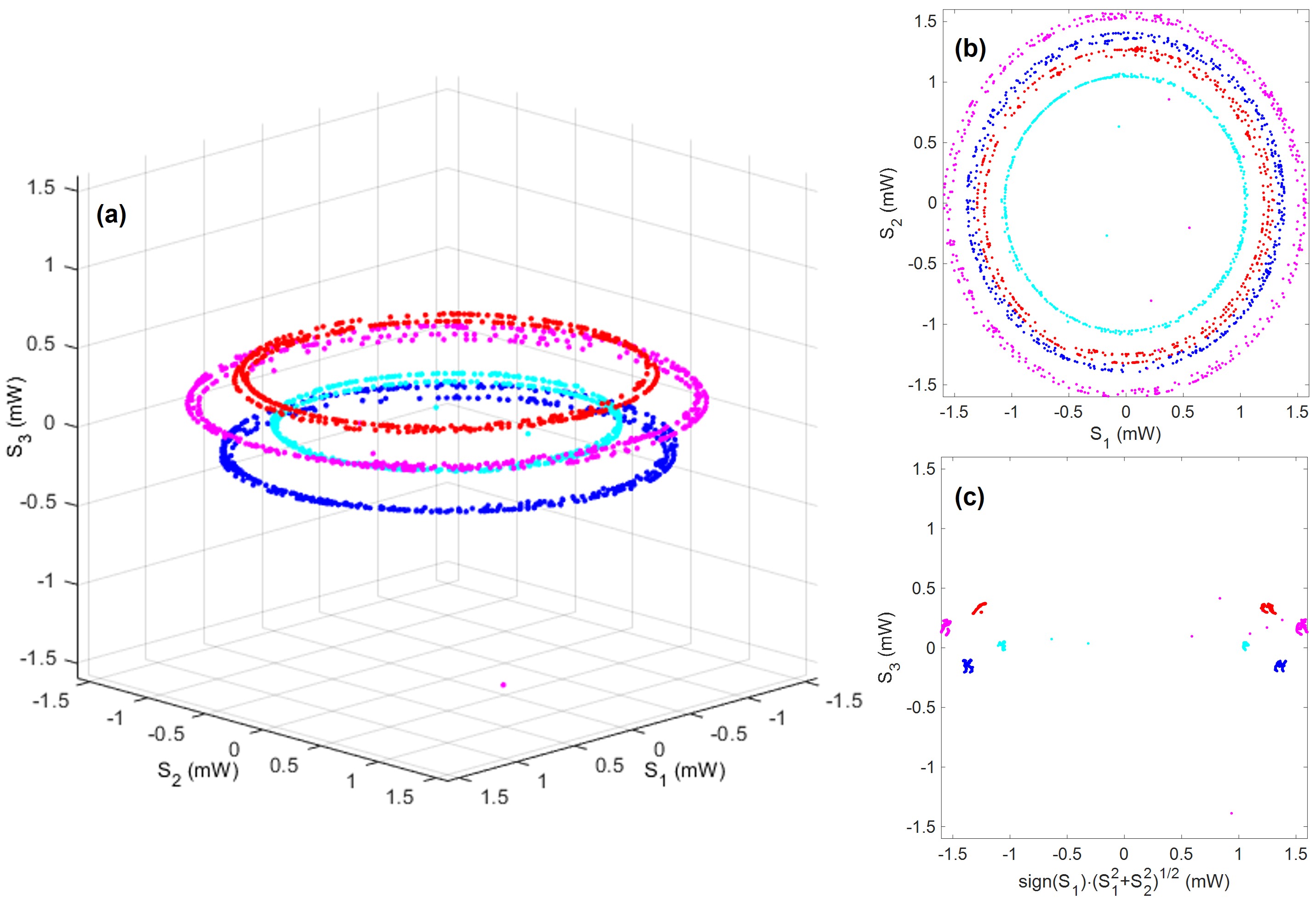}
\caption{
Polarisation torus in the Stokes space.
Stokes parameters for coherent photons out of the polarisation interferometer were plotted.
After setting the poloidal rotation angle, four trajectories (red, magenta, blue, and cyan in colours) were obtained at each angle by mechanically rotating the half-wave-plate to change the toroidal angle. 
Stokes parameters are shown (a) in the 3D Stokes space, (b) from the top of the $S_3$ axis, and (c) as a cross section, perpendicular to the toroidal plane.
}
\end{center}
\end{figure}

\twocolumngrid

\section{Experimental Results}

Our experiments were conducted in a fibre based system, together with short-distance free space optics, as shown in Fig. 2.
A frequency-locked distributed-feedback (DFB) laser-diode (LD) with the wavelength of 1533nm was used and coupled to a SMF.
The polarisation of the input wave was adjusted to the be D-state by a birefringent polarisation-controller (PC), and the input power was 1.8mW.
Then, the input wave was inserted to the polarisation interferometer, as explained above, with 2 FFSs and rotating optical plates, and the output from the interferometer was further controlled by the next rotating optical plates.
At each steps, the polarisation states of the fibre were adjusted by PCs, and the final output wave was examined by a polarimeter to observe the Stokes parameters \cite{Saito22g}.

\subsection{Observation of polarisation torus}

The experimental Stokes parameters are shown in Fig. 6.
We have set $\delta \phi_{\rm p}$ at $0$, $\pi$, $2\pi$, and $3\pi$, which corresponds to the rotation of HWP $\delta {\it \Psi}_{\rm p}$ at $0$, $\pi/4$, $\pi/2$, and $3\pi/4$, respectively.
At each $\delta \phi_{\rm p}$, we have mechanically rotated the HWP of the rotator to change $\delta \phi_{\rm t}$, and obtained the trajectories by recording the Stokes parameters using PM, while rotating physically.
As expected from the proper rotator operation of 2 successive operations of HWPs to form $SO(2)$ \cite{Saito22g}, each trajectory is a circle, located paralell to the $S_1$-$S_2$ plane (Fig. 6 (b)).
On the other hand, the change in $\delta \phi_{\rm p}$ corresponds to the rotation along the poloidal direction, and the intensity has been changed upon the interference, which is confirmed by the small empty region in the cross section of the torus (Fig.6 (c)).
The maximum output power was 1.5mW, such that the minimum insertion loss was about 0.8dB, and the overall feature of the observed polarisation torus is in reasonable agreement with the calculated results (Fig. 3).

Now, we can explain more details on the realisation of the polarisation torus based on the $U(2)$ theory along with our experimental preparations of HWPs and QWPs.
We explain the free space operations in the polarisation interferometer of Fig. 2.
We specify the alignment of these waveplates by the angle of the fast axis (FA), measured from horizon seen from the detector side (opposite to the LD source side) of the plates \cite{Saito20a,Saito22g}.
In our convention, we assume that the angle is 0, if FA is aligned horizontally, equivalently, the slow axis (SA) is aligned vertically, and the direction of rotation is positive, if it is anti-clock-wise rotation, seen from the top of the detector side.
The first free space operations were conducted by a sequential application of QWP (whose FA is aligned to $-pi/4$), HWP (aligned horizontally), HWP (rotated $\delta {\it \Psi}_{\rm p}$), and finally QWP (whose FA is aligned to $pi/4$), and these operations are given by 
\begin{eqnarray}
|{\rm Port \ 4^{\prime}} \rangle 
=&&
\hat{\mathcal{D}} ({\bf \hat{n}_2},\pi/2) 
\hat{\mathcal{D}} ({\bf \hat{n}_3},2\delta {\it \Psi}_{\rm p}) 
\hat{\mathcal{D}} ({\bf \hat{n}_1},\pi) 
\hat{\mathcal{D}} ({\bf \hat{n}_3},-2\delta {\it \Psi}_{\rm p}) 
\nonumber \\
&&
\hat{\mathcal{D}} ({\bf \hat{n}_1},\pi) 
\hat{\mathcal{D}} ({\bf \hat{n}_2},-\pi/2) 
|{\rm Port \ 4} \rangle 
,
\end{eqnarray} 
whose operations are equivalent to $\hat{\mathcal{D}} ({\bf \hat{n}_1},\delta \phi_{\rm p}) $, and the physical rotation of $\pi$ for $\delta {\it \Psi}_{\rm p}$ is enough to realise the equivalent rotation of $4\pi$ for $\delta \phi_{\rm p}$ along the poloidal direction. 

Similarly, the second free space operation can be decomposed into 2 sequential operations of HWP (aligned horizontally) and HWP (rotated $\delta {\it \Psi}_{\rm t}$)
\begin{eqnarray}
|{\rm Output} \rangle 
&=&
\hat{\mathcal{D}} ({\bf \hat{n}_3},2\delta {\it \Psi}_{\rm t}) 
\hat{\mathcal{D}} ({\bf \hat{n}_1},\pi) 
\hat{\mathcal{D}} ({\bf \hat{n}_3},-2\delta {\it \Psi}_{\rm t}) 
\nonumber \\
&&
\hat{\mathcal{D}} ({\bf \hat{n}_1},\pi) 
|{\rm Port \ 1}^{\prime} \rangle 
,
\end{eqnarray}
to realise the proper $SO(2)$ rotation \cite{Saito22g} by $\hat{\mathcal{D}} ({\bf \hat{n}_3},\delta \phi_{\rm t})$, and the physical rotation of $\pi/2$ for $\delta {\it \Psi}_{\rm t}$ is enough to realise the equivalent rotation of $2\pi$ for $\delta \phi_{\rm t}$ along the toroidal direction.

\onecolumngrid

\begin{figure}[h]
\begin{center}
\includegraphics[width=16cm]{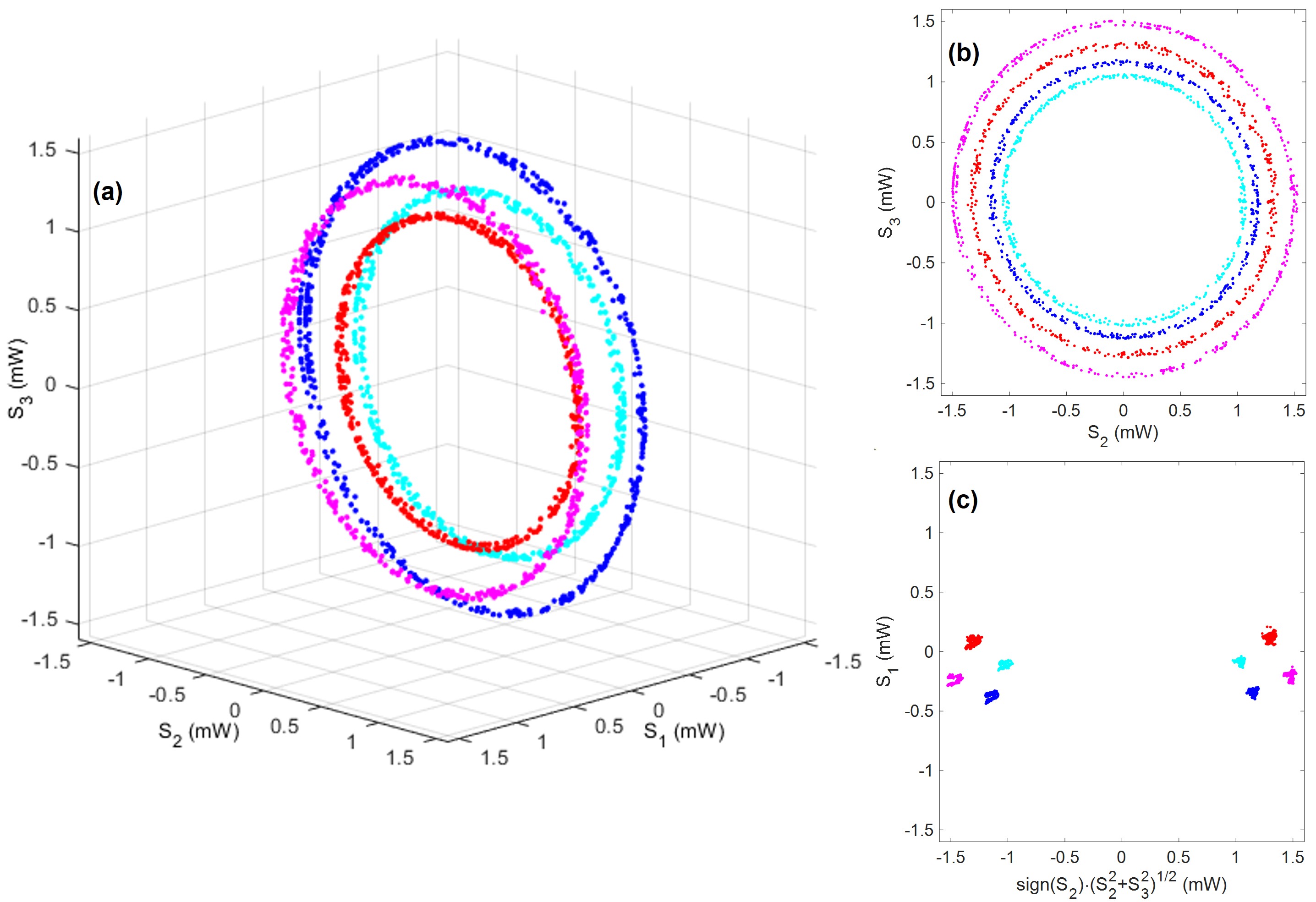}
\caption{
Rotated polarisation torus.
The quarter-wave-plate, whose fast axis was aligned to the diagonal direction, was inserted.
Stokes parameters are shown (a) in the 3D Stokes space, (b) from the top of the $S_3$ axis, and (c) as a cross section, perpendicular to the toroidal plane.
We confirmed that the torus was rotated $\pi/2$, such that the toroidal direction is parallel to the $S_2$-$S_3$ plane.
}
\end{center}
\end{figure}

\twocolumngrid

%\clearpage

\subsection{Rotated polarisation torus}

A torus is obviously a topological structure, such that it is expected that the topological structure is strong against distortions.
As the first step to confirm the topological robustness, we have inserted the additional QWP with its FA rotated $\pi/4$ towards the end of the device, such that the output becomes
\begin{eqnarray}
|{\rm Port \ 4^{\prime}} \rangle 
&=&
\hat{\mathcal{D}} ({\bf \hat{n}_2},\pi/2) 
\hat{\mathcal{D}} ({\bf \hat{n}_1},\delta \phi_{\rm p}) 
|{\rm Port \ 4} \rangle 
,
\end{eqnarray}
which means that the polarisation states should be rotated along the $S_2$ axis with the amount of $\pi/2$, and consequently, the $S_1$-$S_2$ plane is rotated to be the $S_2$-$S_3$ plane.
The experimental results on the rotated torus is shown in Fig. 7.
We confirmed that the structure of the torus remained unchanged in the Stokes space upon the rotation, while the toroidal direction is now located parallel to the $S_2$-$S_3$ plane.

Similarly, we have also confirmed that the rotation along the $S_1$ axis preserves the topological structure of the torus.
This was realised by adding QWP with its FA aligned horizontally, and the expected output state becomes 
\begin{eqnarray}
|{\rm Port \ 4^{\prime}} \rangle 
&=&
\hat{\mathcal{D}} ({\bf \hat{n}_1},\pi/2) 
\hat{\mathcal{D}} ({\bf \hat{n}_1},\delta \phi_{\rm p}) 
|{\rm Port \ 4} \rangle 
,
\end{eqnarray}
and the experimental results are show in Fig. 8.
In this case, the principal axis of the toroidal rotation is along the $S_2$ axis, which is orthogonal to the axes of previous toruses (Figs. 6 and 7).
Therefore, the torus was not distorted upon the applications of rotated QWPs.

Theoretically, the application of the phase-shifter and the rotator merely rotate Stokes parameters upon the application of $\hat{\mathcal{D}} ({\bf \hat{n}},\delta \phi) $, such that it results in rotations of the vectorial point $(S_1,S_2,S_3)$ along some direction ${\bf \hat{n}}$ with the amount of $\delta \phi$.
This linear and unitary operation cannot change the topology of a set of points in the Stokes space, such that a torus or a sphere would be transferred to the same topological structure, respectively, without changing its genus.

\onecolumngrid

\begin{figure}[h]
\begin{center}
\includegraphics[width=16cm]{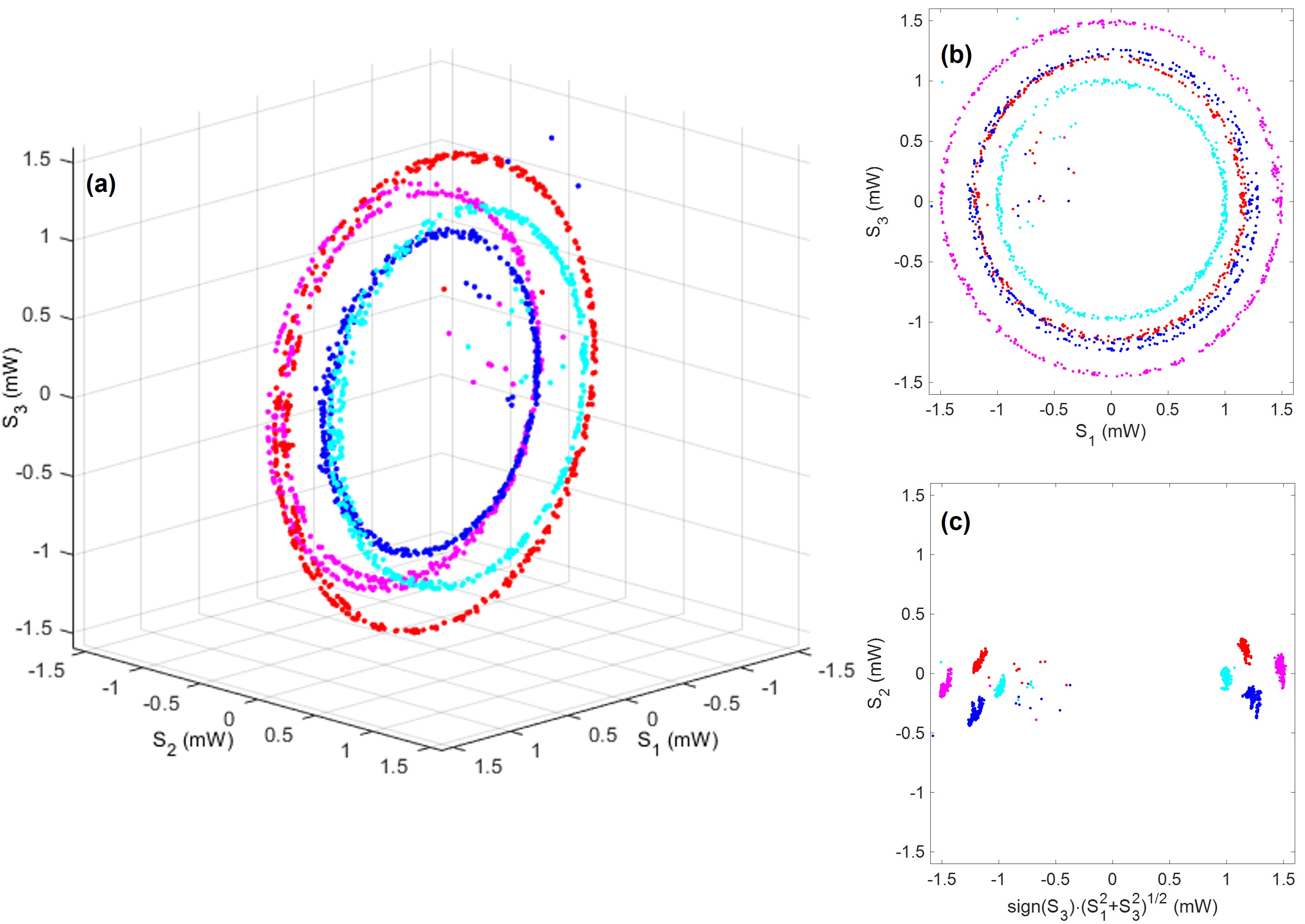}
\caption{
Rotated polarisation torus.
The quarter-wave-plate, whose fast axis was aligned horizontally, was inserted.
Stokes parameters are shown (a) in the 3D Stokes space, (b) from the top of the $S_3$ axis, and (c) as a cross section, perpendicular to the toroidal plane.
We confirmed that the topological structure of the torus was not changed upon the rotation.
}
\end{center}
\end{figure}

\twocolumngrid

%\clearpage

\subsection{Double connected toruses}

\begin{figure}[h]
\begin{center}
\includegraphics[width=8cm]{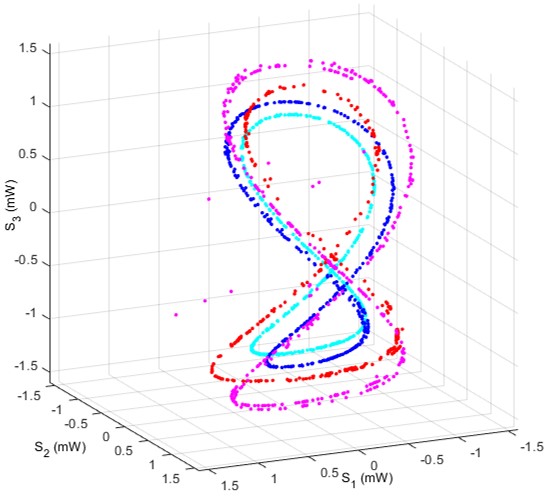}
\caption{
Double connected toruses.
The polarisation states were made by operating the states by using a rotated quarter-wave-plate at the end of the polarisation interferometer.
The sets are made of 2 connected toruses with genus of 1.
}
\end{center}
\end{figure}

We have also tried to realise the double connected toruses with $g=1$ by a rotated QWP.
Here, we expect the output state of 
\begin{eqnarray}
|{\rm Output} \rangle 
&=&
\hat{\mathcal{D}} ({\bf \hat{n}_3},2\delta {\it \Psi}_{\rm t}) 
\hat{\mathcal{D}} ({\bf \hat{n}_1},\pi/2) 
\hat{\mathcal{D}} ({\bf \hat{n}_3},-2\delta {\it \Psi}_{\rm t}) 
|{\rm Port \ 1}^{\prime} \rangle 
,
\nonumber \\
\end{eqnarray}
which makes 2 connected toruses.
The corresponding experimental results are shown in Fig. 9.
Each trajectories are connected near the D-state, such that there is no hole around this region to unify the whole structure as a unit manifold.
Therefore, the set of points cannot be the torus of $g=2$, and instead, the set is separated into two sets of toruses with $g=1$.

%\clearpage

\section{Discussions}

We have shown polarisation torus is realised in the Stokes space, where each point represents spin expectation value of photon coherent states.
Next, we consider other examples of topologically non-trivial polarisation states.

%\clearpage

\subsection{M\"{o}bius strip}

We consider how to realise a M\"{o}bius strip in the Stokes space.
A M\"{o}bius strip \cite{Tanda02,Cisowski22,Nakahara90} is made of a strip with the one end, flipped for connecting to the other end.
We consider the same set-up with that shown in Fig. 2, and we just need to change the optical components in the free space regions.
For a M\"{o}bius strip, we need to prepare a line segment, rather than a circle (Fig. 1), prepared for the torus.
Such a line segment can be made by the simple phase-shift to the wave out of the tap port 4 by the phase-shift of $\hat{\mathcal{D}} ({\bf \hat{n}_2},\delta \phi_{\rm p}) $ along the $S_2$ axis.
This operation will not change the input polarisation state of the D-state, while the phase is shifted, which changes the intensity of $S_0$.
The line segment should be rotated for the toroidal direction along the $S_1$ axis, such that the operations become 
\begin{eqnarray}
|{\rm Port \ 4^{\prime}} \rangle 
=
\hat{\mathcal{D}} ({\bf \hat{n}_1},\delta \phi_{\rm t}) 
\hat{\mathcal{D}} ({\bf \hat{n}_2},\delta \phi_{\rm p}) 
|{\rm Port \ 4} \rangle 
. 
\end{eqnarray}
After these operations, the bypassed wave should be recombined by the subsequent FFC with the output from the through port 3.
The combined wave should be rotated by the final rotator with the amount of $\delta \phi_{\rm t}$, which is the same operation with that for the torus.
The Stokes parameters were calculated from these $U(2)$ wavefunctions, and the trajectories become the polarisation M\"{o}bius strip, as shown in Fig. 10 (a).
We can recognise a standard feature of a M\"{o}bius strip, designed in the Stokes space.

In our original consideration for the torus, we have controlled $\delta \phi_{\rm p}$ from 0 to $4\pi$, while $\delta \phi_{\rm t}$ was changed from 0 to $2\pi$.
These parameters also work for the M\"{o}bius strip, this will cover the M\"{o}bius strip twice, due to the collapsing of the pore of the torus for the strip.
Consequently, the half-rotation for $\delta \phi_{\rm p}$ is enough to realise the M\"{o}bius strip.
It is even better to consider the quarter-rotation of $\delta \phi_{\rm p}$ from 0 to $\pi$, and instead, the twice-rotation of $\delta \phi_{\rm t}$ from 0 to $4\pi$ could be considered.
In this case, we it is easier to track the trajectory, controlled by $\delta \phi_{\rm t}$.
For example, if we start from the point, realised by $\delta \phi_{\rm p}=0$ and  $\delta \phi_{\rm t}=0$, the original input state is recovered, which is located at the edge of the M\"{o}bius strip and the intensity is maximised.
Then, it is easy to see the trajectory (Fig. 10 (b)) of showing how $\delta \phi_{\rm t}$ changes the point, moving from the outer edge to the inner edge, and coming back to the original point upon the application of the $4\pi$-rotation.
On the other hand, if we start from $\delta \phi_{\rm p}=\pi$ and  $\delta \phi_{\rm t}=0$, it is located at the centre of the M\"{o}bius strip, and the trajectory becomes the circle upon the change of  $\delta \phi_{\rm t}$ from $0$ to $2\pi$, and it rotates twice upon the $4\pi$-rotation (Fig. 10 (c)).

\onecolumngrid

\begin{figure}[h]
\begin{center}
\includegraphics[width=12cm]{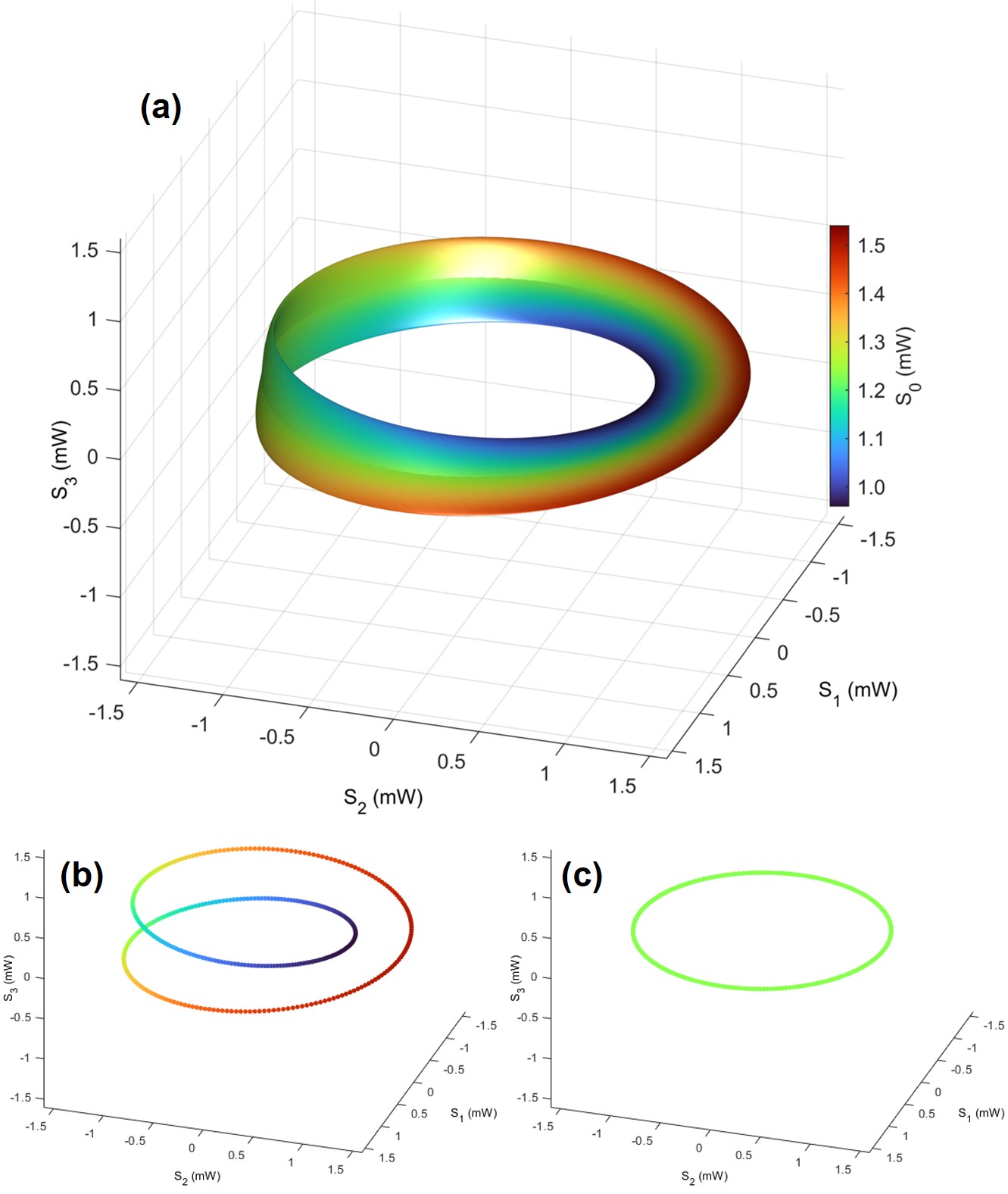}
\caption{
Polarisation M\"{o}bius strip.
Stokes parameters were calculated, assuming the input of 1.5mW to a polarisation interferometer.
A line segment is made of the interference between the bypassed wave with the phase-shift and the original polarisation state at the through port.
The segment was rotated by a rotator for the same amount of rotation with the angle for the toroidal rotation.
This M\"{o}bius strip is right-handed, in the sense that it is made of the line segment, which rotates to the right, seen from the direction, opposite to the toroidal rotation.
(a) Set of polarisation states in the Stokes space, realised by the polarisation interferometer and rotators. 
(b) Swapping of the edges of the M\"{o}bius strip. A trajectory of polarisation states at the poloidal angle of $\delta \phi_{\rm p}$=0 is shown, while the toroidal angle of  $\delta \phi_{\rm t}$ was changed from $0$ to $4\pi$. 
The outer edge state becomes the inner edge state, after 1-rotation, and {\it vice versa}, after subsequent another rotation.
(c) The centre line of the M\"{o}bius strip. A trajectory of polarisation states at the poloidal angle of $\delta \phi_{\rm p}=\pi$ becomes a polarisation circle, to keep the centre of the M\"{o}bius strip, while rotating.
}
\end{center}
\end{figure}

\twocolumngrid

\clearpage

It is well-known that a M\"{o}bius strip cannot be assigned its orientation, which is evident from the fact that we cannot distinguish the front surface with the back surface.
On the other hand, we can define its chirality, which depends on how the line segment could be connected upon rotations in our case.
The M\"{o}bius strip, as shown in Fig. 10, is defined to be right-handed, because the line segment was rotate to the right side, seen from the direction, opposite to the toroidal rotation.
Correspondingly, the left-handed M\"{o}bius strip could be considered, simply by the opposite rotation to yield,
\begin{eqnarray}
|{\rm Port \ 4^{\prime}} \rangle 
=
\hat{\mathcal{D}} ({\bf \hat{n}_1},\delta \phi_{\rm t}) 
\hat{\mathcal{D}} ({\bf \hat{n}_2},-\delta \phi_{\rm p}) 
|{\rm Port \ 4} \rangle 
, 
\end{eqnarray}
which was used to calculate its Stokes parameters, as shown in Fig. 11.
If we focus on the trajectory of the edge, starting from the outer edge at $\delta \phi_{\rm p}=0$ and  $\delta \phi_{\rm t}=0$, polarisation states rotate along the bottom (negative $S_3$), upon the toroidal direction, to arrive at the inner edge, and continue to go up to the top (positive $S_3$), towards going back to the original point.
This is the opposite chirality to that of right-handed M\"{o}bius strip, shown in Fig. 10.
Therefore, the chirality could be controlled, when we realise the M\"{o}bius strip by defining the sign of rotation to flip the line segment.

%\clearpage

\begin{figure}[h]
\begin{center}
\includegraphics[width=8cm]{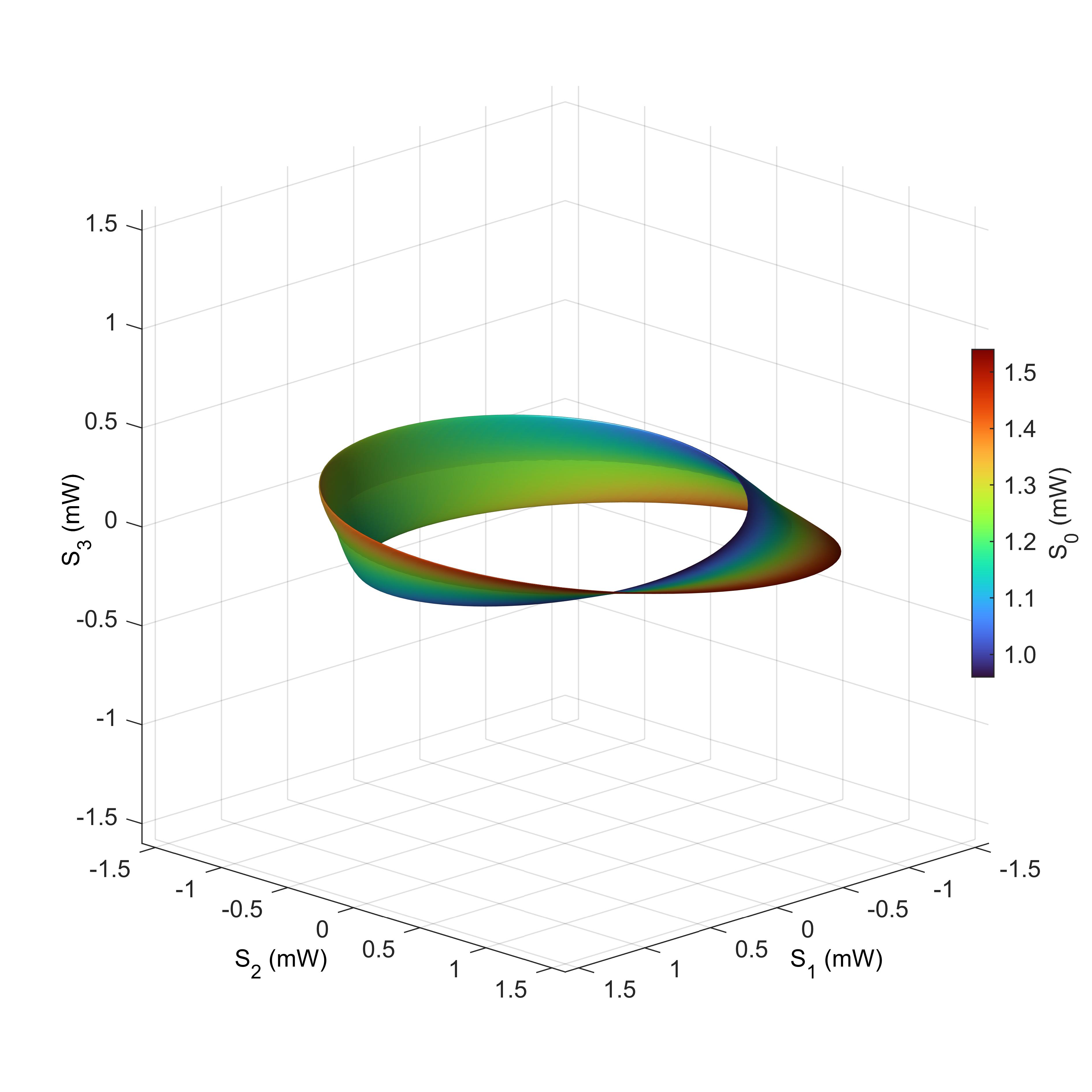}
\caption{
Left M\"{o}bius strip as polarisation states. 
Stokes parameters were calculated by considering the rotating segment to the left-handed direction, seen from the direction, opposite to the toroidal rotation.
}
\end{center}
\end{figure}

%\clearpage

\subsection{Hopf-links and other topological structures}

Next, we consider to realise a Hopf-link using polarisation states.
A Hopf link is made of two rings, which are completely disconnected, while one ring is intersecting to the other (Figs. 12 and 13).
We think it is impossible to realise it solely from 1 wavelength, since we cannot allow 2 different points with different $N$ (or equivalently, $P$), while keeping the same angles for $\gamma$ and $\delta$.
Therefore, trajectories, controlled by the polarisation states, would be continuous in the Stokes space for 1 wavelength.
However, if we allow wavelength-division-multiplexing (WDM) in the SMF, we can separately manipulate polarisation states for multiple wavelengths, leading the way to realise an optical Hopf-links.
We just need to adjust relative powers and polarisation states for both wavelengths by considering to establish topology between 2 waves.

As an example, we consider a polarisation circle realised by the polarisation interferometer, shown in Fig. 2.
Here, we assume 1 wavelength of say at 1530nm with the input power of 1.5mW to be controlled by the polarisation interferometer.
The output of the tap port 4 is now controlled as 
\begin{eqnarray}
|{\rm Port \ 4^{\prime}} \rangle 
&=&
\hat{\mathcal{D}} ({\bf \hat{n}_3},\delta \phi_{\rm p}) 
|{\rm Port \ 4} \rangle 
, 
\end{eqnarray}
while we do not need to rotate along the toroidal direction, such that we do not insert any optical components in the second free space region.
This will create a polarisation circle in the $S_1$-$S_2$ plane, perpendicular to the direction of $S_3$ (the small circle of Fig. 12).
The polarisation circle in the Stokes space is a line segment in the normalised Poincar\'e sphere (the blue line in the inset of Fig. 12).
Next, we just need to prepare another polarisation circle by using a different wavelength of say 1550nm at the power of 1.0mW, which could be controlled by proposed Poincar\'e rotators  (phase-shifters to control $\delta$) \cite{Saito22g,Saito22h}, separately, to allow circular changes of polarisation states, located in the $S_2$-$S_3$ plane, perpendicular to the direction of $S_1$ (a large circle of Fig. 12). 
After constructing these waves in the SMFs, we can combine these by a polarisation dependent FFC, with appropriate power splitting ratio
These 2 polarisation circles are not touching each other in Stokes space, forming a Hopf-link (Fig. 12).
On the other hand, if we plot these states by normalising Stokes parameters to have unit radius, these 2 circles are connected (the inset of Fig. 12).
Therefore, it is important to distinguish the power difference of these waves.

\begin{figure}[h]
\begin{center}
\includegraphics[width=8cm]{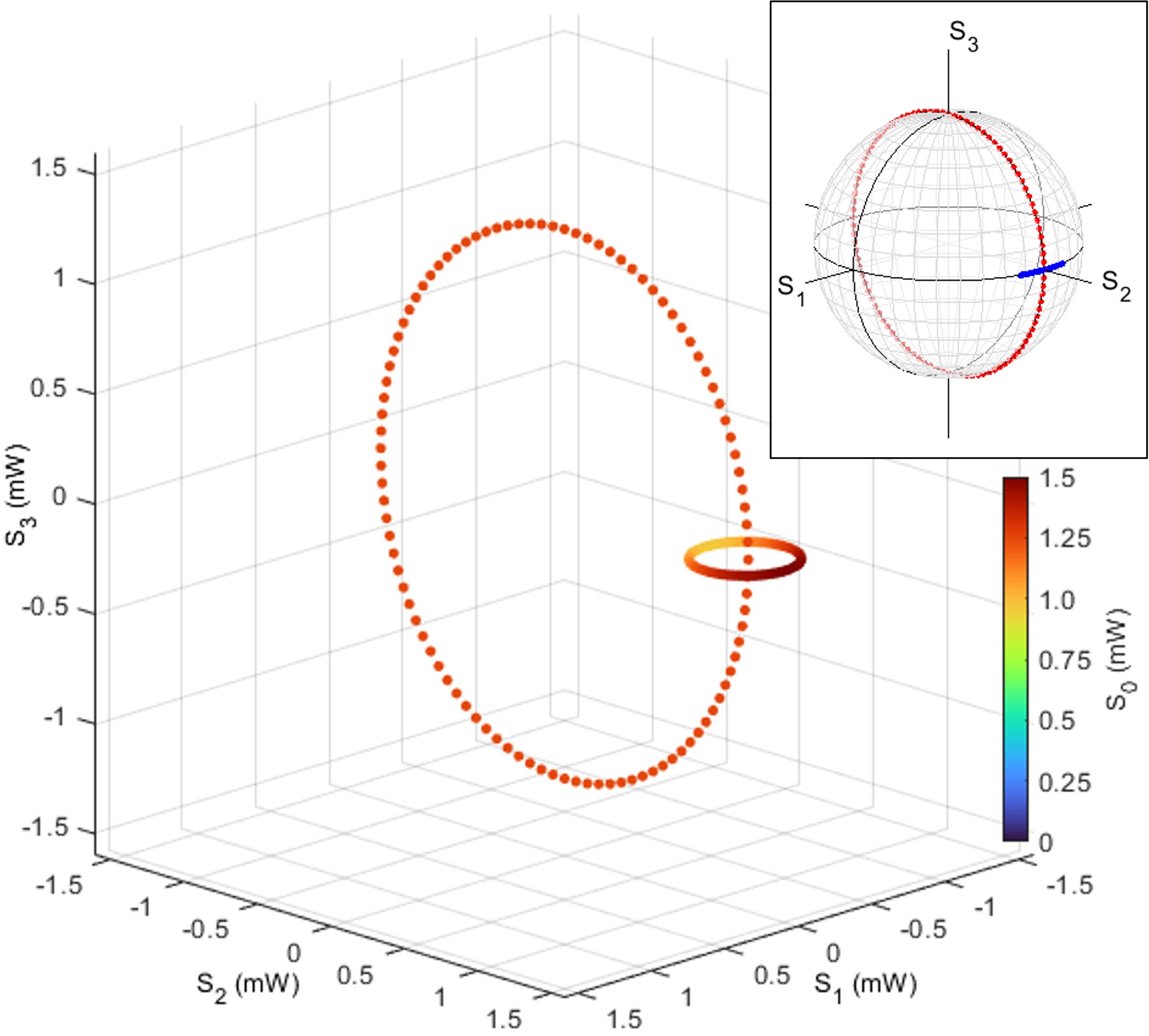}
\caption{
Polarisation Hopf link in the Stokes space.
The small circle is calculated for the wavelength of 1530nm, realised by the polarisation interferometer.
The large circle is realised by a phase-shifter for the wavelength of 1550nm, and these 2 waves would be combined by a coupler.
The inset shows the normalised polarisation states, shown on the Poincar\'e sphere. The blue (red) circles are for small (large) polarisation circles.
}
\end{center}
\end{figure}

Similarly, we have also calculated a Hopf-link by assuming the different ratio of 50:50 ($\alpha=0.5$), as shown in Fig. 13, while the other parameters were the same as those for Fig. 12.
In this case, the larger polarisation circle is realised upon the interference, due to the larger power, propagating into the tap port 4.
As a consequence of the interference, the minimum output power could be zero, that is why the heart-like dip is realised near the origin of the Stokes space.

\begin{figure}[h]
\begin{center}
\includegraphics[width=8cm]{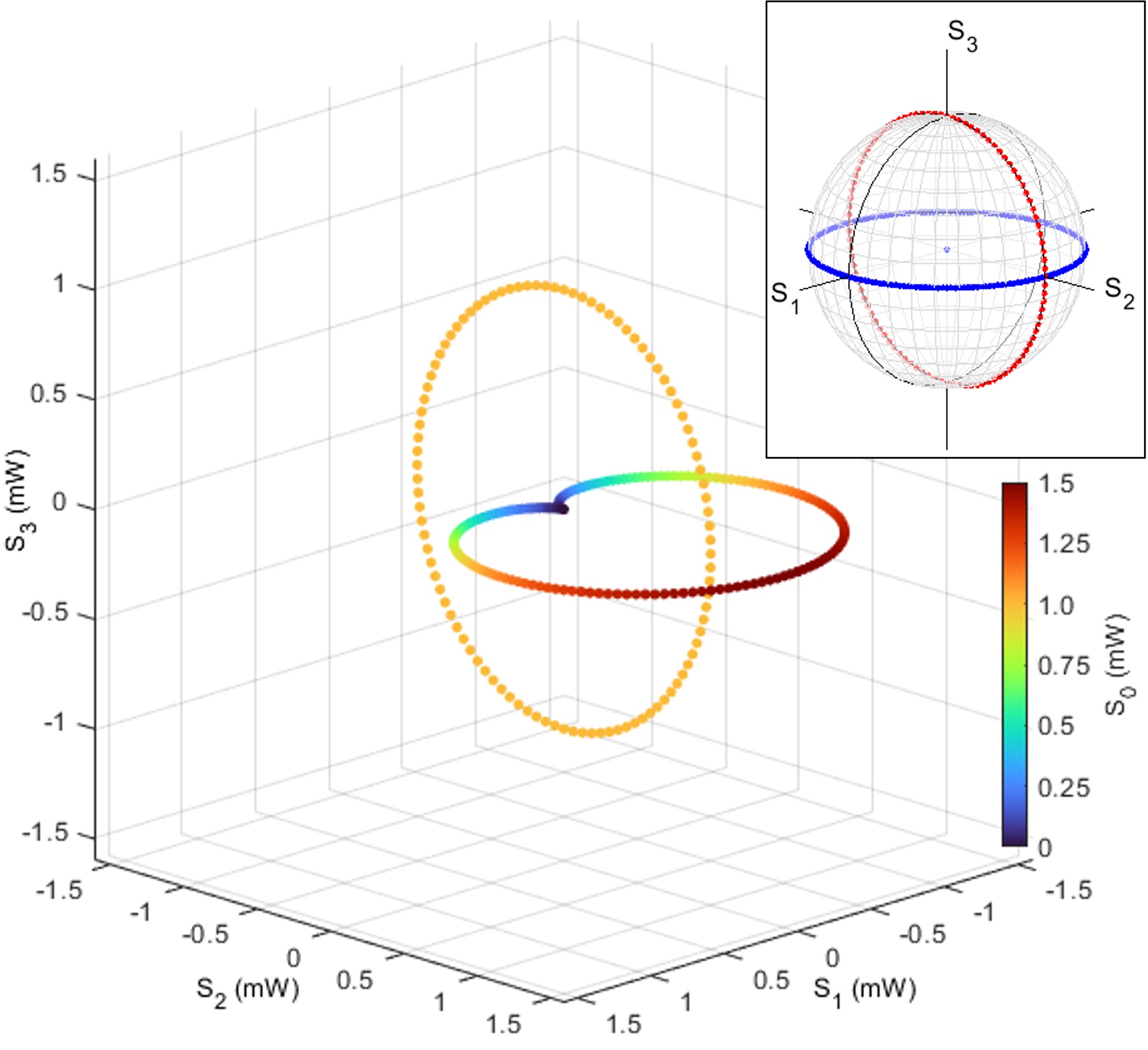}
\caption{
Heart-shaped polarisation circle with a Hopf link in the Stokes space.
We assumed 50:50 splitting of the input wave into the polarisation interferometer to realise the hart-like dip near the origin.
The inset shows the normalised polarisation states, shown on the Poincar\'e sphere. 
The normalised polarisation circle (the blue line in the inset) cover the whole angle in the $S_1$-$S_2$ plane, while the intensity is modulated upon rotations.
}
\end{center}
\end{figure}

This heart-shape affects the torus structure, if the polarisation states are further controlled upon the toroidal rotation, as shown in Fig. 14.
Here, we have assumed the splitting of 50:50 ($\alpha=0.5$) for an input of the single wavelength at 1530nm with the power of 1.5mW.
We expect the the heart-shape polarisation circles are making trajectories upon rotations to the toroidal direction with the amount of $\delta \phi_{\rm t}$, changed from $0$ to $3\pi/2$.
Due to its feature, we call it as a polarisation apple to have seeds-like regions due to the heart-dip near the origin. 
This is a remarkable difference between mathematical overlapping of 2 circles upon rotations.
In our case, we realise the circular heart-shaped circles upon the interference, such that the intensity near the origin becomes zero due to the complete destructive interference between 2 separated waves with the sign change upon the $SU(2)$ rotation.

\begin{figure}[h]
\begin{center}
\includegraphics[width=8cm]{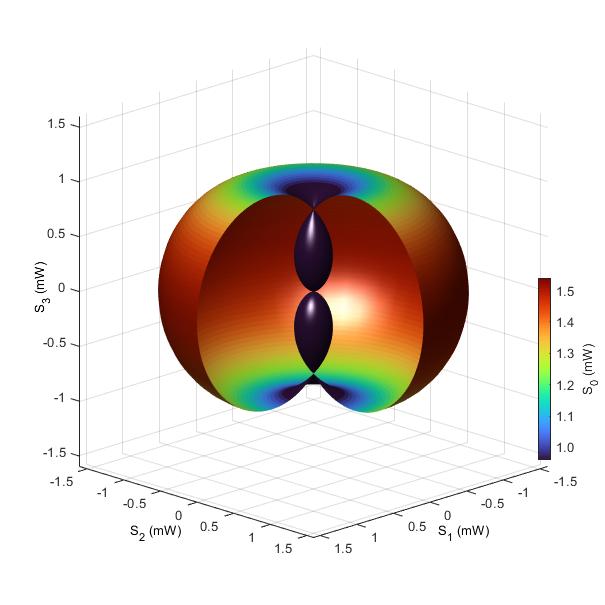}
\caption{
Polarisation apple in the Stokes space.
The stokes parameters were calculated for assuming 50:50 splitting at fibre couplers.
The toroidal rotations from $0$ to $3\pi/2$ (rather than $2\pi$) were assumed to see the seeds-like dips near the origin, realised by the interferences.
}
\end{center}
\end{figure}

%\clearpage

\subsection{Bulk-edge correspondence and massless Dirac bosons}

We have several topologically non-trivial structures as polarisation states in the Stokes space, compared with the Poincar\'e sphere.
Polarisation is coming from spin expectation values \cite{Stokes51,Poincare92,Max99,Jackson99,Yariv97,Gil16,Goldstein11,Parker05,Chuang09,Hecht17,Pedrotti07,Grynberg10,Jones41,Fano54,Baym69,Sakurai67,Sakurai14,Saito20a,Saito20c,Saito22g,Saito22h}, such that the non-trivial polarisation states are determined by the broken rotational symmetry for the polarisation states.
These topologically non-trivial features are robust against the rotationally symmetric disturbances. 
For examples, the polarisation independent loss in the SMF can not change the topology of the polarisation states.
Moreover, spherically symmetric operations of phase-shifters and rotators can rotate the topological structures such as toruses (Figs. 6-8), M\"{o}bius strip (Figs. 10 and 11), and Hopf-links (Figs. 12 and 13), will be rotated, but still the relative topology within these structures will be kept upon rotations.
It is a polarisation dependent loss to cut these topological features.
Nevertheless, it is not so easy to change the topology, since a simple insertion of a polariser, for example, will completely destruct the polarisation structure, ending up to be one point in the Stokes space.
We need to make an optical {\it scissor} to allow an arbitral cutting of topological polarisation states.
In order to reduce the intensity of the targetted polarisation states, we need to observe the polarisation states by using a polarimeter, which corresponds to observe the wavefunctions to expect the collapse of the wavefunction.
For coherent photons, we can observe a bypassed contribution via a tap port, while keeping the contribution in the through-port \cite{Saito22h}, but still we need to prepare a complicated photonic circuit to allow the splitting, delay, and manipulations of the loss.
These difficulties are coming from robust correlations among the bits in the pulse stream to form a non-trivial topological polarisation states.
Here, we discuss how the broken symmetry in the bulk is corresponding to the edge state \cite{Laughlin81,Thouless82,Kohmoto85,Hatsugai93,Moore10,Hasan10,Qi11,Nakahara90,Cisowski22}.
More specifically, we go back to the case of the polarisation torus, and consider how the polarisation states could be connected to the original Poincar\'e sphere.

The torus is obviously distinct with the sphere, because of the non-zero genus.
For the pulse streams of light, coming out of the proposed device, which is the polarisation interferometer with the polarisation rotator (Fig. 2), the Stokes parameters of the pulse represent one of the points on the polarisation torus, such that the pulse streams are considered to form the bulk state of the polarisation torus, characterised by $g=1$.
On the other hand, in the standard SMF with the rotational symmetry, the polarisation states are well known to be characterised by the  Poincar\'e sphere with $g=0$ as a different bulk state.
If we would like to connect the torus to the Poincar\'e sphere, we need to prepare the edge state, where the pore of the torus is closed, such that the polarisation circle (Fig. 1(a)) is closed to be 1 point, which must be robust against disturbances to rotate the polarisation state.

\begin{figure}[h]
\begin{center}
\includegraphics[width=8cm]{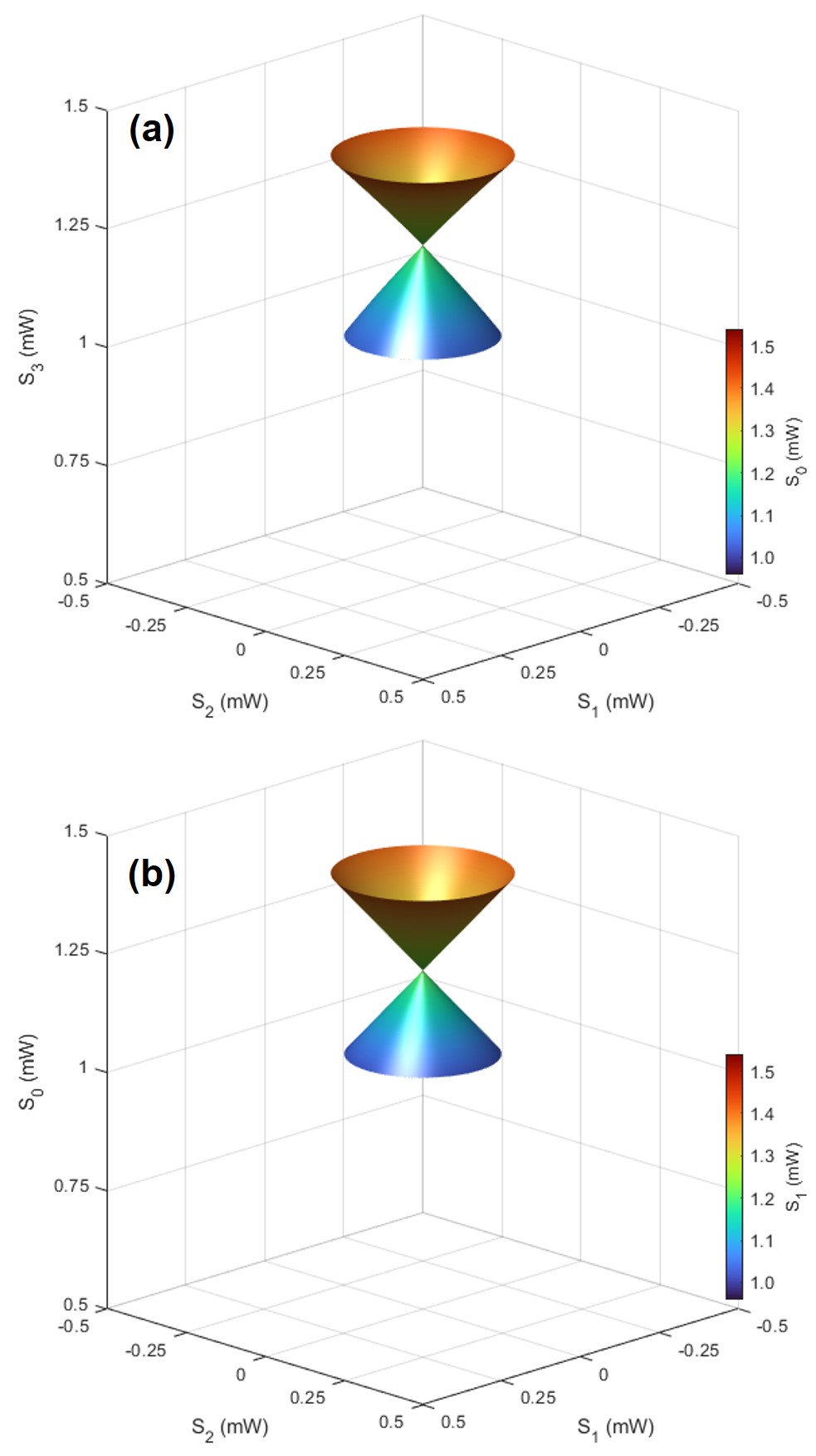}
\caption{
Topological Dirac bosons.
Stokes parameters $(S_0,S_1,S_2,S_3)$ were calculated for the output from the amplitude controlled polarisation interferometer.
(a) Dirac bosons in the vectorial space $(S_1,S_2,S_3)$. 
The Dirac point at the centre of the light cone is robust against the rotation in the $S_1$-$S_2$ plane, such that a rotator along the $S_3$ axis cannot change the polarisation state at this point.
(b) Dirac bosons in the vectorial space $(S_0,S_1,S_2)$. 
The intensity of the light ($S_0$) is linearly controlled upon the $S_3$ direction, as a result of the interference, while $S_0$ is not affected upon the rotation in the $S_1$-$S_2$ plane, which induces a helical change of the polarisation state, 
}
\end{center}
\end{figure}

We can generate such an edge state, simply by changing the rotations in the Stokes space.
One practical implementation is to use the same set-up of Fig. 2, while we introduce the amplitude controller,
\begin{eqnarray}
\hat{\mathcal{A}} (\delta \phi_{\rm p}) 
&=&
{\bf 1}
\cos
\left (
\frac{\delta \phi_{\rm p}}{2}
\right),
\end{eqnarray}
in the first free space region.
The amplitude controller is not a unitary operator, such that this operator does not belong to a family of $SU(2)$.
It is an operator of $U(2)$, except for the parameters, $\delta \phi_{\rm p}=\pi, 3\pi, \cdots$, where the operator becomes ${\bf 0}$, such that the norm is controlled upon the operation.
The amplitude controller is made of polarisation splitters to change the amplitude of each polarisation states, independently, while the each polarisation state would be inserted into a polarisation rotator \cite{Saito20a,Saito22g,Saito22h} to change its polarisation in the $S_1$-$S_2$ plane, independently, picking up only the original polarisation state after the rotation, and finally recombining orthogonal polarisation states in a combiner.
Alternatively, the amplitude controller is simply made of a polarisation independent Mach-Zehnder interferometer to control the amplitudes for both polarisation components, simultaneously, while keeping the polarisation.
The amplitude controller could also be defined as 
\begin{eqnarray}
\hat{\mathcal{B}} (\delta \phi_{\rm p}) 
&=&
\hat{\sigma}_1
\sin
\left (
\frac{\delta \phi_{\rm p}}{2}
\right)
, 
\end{eqnarray}
which accompanies a swapping of the polarisation states in addition to the polarisation rotation.

After the amplitude control of the contribution in the tap port 4, the output polarisation state is further controlled by a QWP, yielding 
\begin{eqnarray}
|{\rm Port \ 4^{\prime}} \rangle 
=
\hat{\mathcal{D}} ({\bf \hat{n}_1},\pi/2) 
\hat{\mathcal{A}} (\delta \phi_{\rm t}) 
|{\rm Port \ 4} \rangle 
, 
\end{eqnarray}
which is combined with the contribution with through port 3.
Finally, the recombined state is rotated along the $S_2$ axis instead of the $S_3$ axis by the polarisation rotator
\begin{eqnarray}
|{\rm Output} \rangle 
&=&
\hat{\mathcal{D}} ({\bf \hat{n}_2},\delta \phi_{\rm t}) 
|{\rm Port \ 1}^{\prime} \rangle 
.
\end{eqnarray}
These operations will create a topological Dirac cone near the D-state.
But, it is intriguing to illustrate it near the north pole (in our convention, the left circularly polarised state \cite{Saito20a}), which could be achieved simply by a $\pi/2$-rotation along $S_1$.
Alternatively, we can apply 
\begin{eqnarray}
|{\rm Port \ 4^{\prime}} \rangle 
=
\hat{\mathcal{A}} (\delta \phi_{\rm t}) 
|{\rm Port \ 4} \rangle 
, 
\end{eqnarray}
to the tap port 4 without a QWP, while the through port 3 is phase-shifted by a QWP to be 
\begin{eqnarray}
|{\rm Port \ 3^{\prime}} \rangle 
=
\hat{\mathcal{D}} ({\bf \hat{n}_1},\pi/2) 
|{\rm Port \ 3} \rangle 
, 
\end{eqnarray}
and then, the recombined states should be rotated along the $S_3$ axis, just like creating a torus before.

The calculated Stokes parameters in this way are shown in Fig. 15, where a topological Dirac cone is recognised in the Stokes space.
Bosonic Dirac boson were previously discussed \cite{Kumar20,Banerjee16} in realising as a single particle spectrum of a boson.
Here, we are not discussing a single particle energy spectrum in the momentum space.
In stead, we are considering the many-body energy ($S_0$) of a bit in a pulse stream, generated from a device, and the change in energy is described against the spin expectation values rather than momentum for an energy band.
Due to the coherent nature of bosons with no charge, photons in the same bit are not interacting each other, but the energy difference in $S_0$ could be considered as the difference in number of photons in each bit.
As shown in Fig. 15 (a), the polarisation circle, generated by a rotator, is closed at the Dirac point, where the light cone is closed.
At the Dirac point, the polarisation state is not changed upon rotations along the $S_3$ axis, such that this edge state to close the circle is topologically robust against rotations in the $S_1$-$S_2$ plane.
We have also plotted the Dirac cone in the space for $(S_0,S_1,S_2)$, as shown in Fig. 15 (b). 
We confirmed the linear energy change, seen from $S_0$, against the radius of the polarisation circle in the plane, parallel to the $S_1$-$S_2$ plane.
For the constant energy in $S_0$, or equivalently, for the same number of photons in a bit, the bit is characterised in the polarisation circle of the Dirac cone, which changes the helical spin expectation values $(S_1,S_2$ upon rotations by a rotator along the $S_3$ axis.

If we would like to change $g=1$ of a torus to $g=0$ of a sphere, it is inevitable to make such a Dirac point, where the polarisation state is robust against the rotations.
We think the Dirac point corresponds to the edge state, while the polarisation torus and Poincar\'e sphere are bulk states.
This is the bulk-edge correspondence 
\cite{Laughlin81,Thouless82,Kohmoto85,Hatsugai93,Moore10,Hasan10,Qi11,Nakahara90,Cisowski22} 
for our topological polarisation states.

\begin{figure}[h]
\begin{center}
\includegraphics[width=8cm]{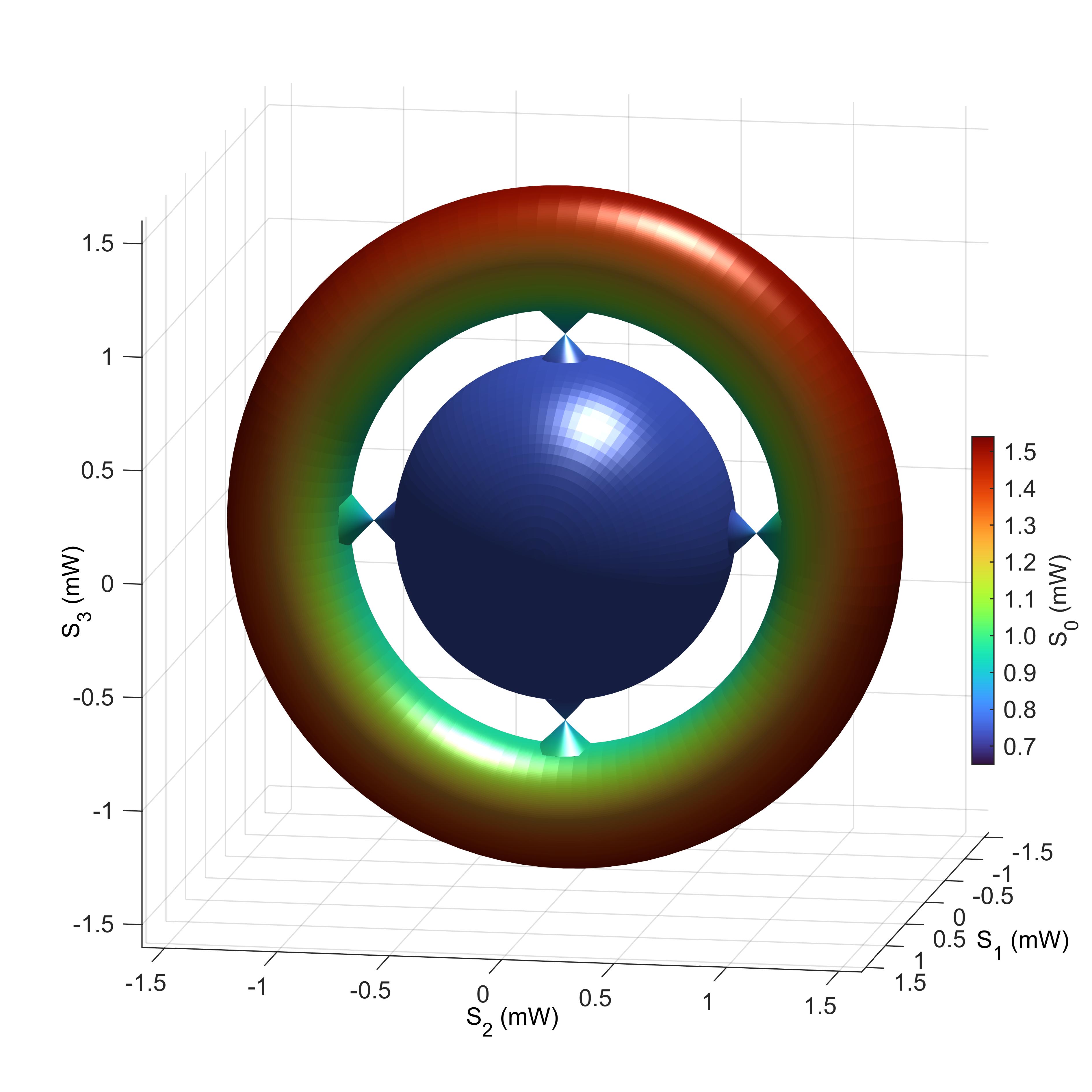}
\caption{
Bulk-edge correspondence for topological polarisation states.
Stokes parameters were calculated for the torus, Dirac cones, and the Poincar\'e sphere.
The polarisation torus was connected by Dirac cones to the Poincar\'e sphere.
It is inevitable to have a node of the polarisation circle at the Dirac point, to change the genus from 1 to 0.
}
\end{center}
\end{figure}

As an example, we have connected a polarisation torus to a Poincar\'e sphere vis topological Dirac cones, as shown in Fig. 17.
Here, we have assumed the maximum power of 1.5mW, 1.05mW, and 0.75mW for the polarisation torus, Dirac bosons, and the inner Poincar\'e sphere, respectively.
The number of connected Dirac cones were 4 in this example, but it is not limited to this particular number.
The number of edge states simply depend on our experimental set-up and feasibility on how to close the pore generate in the torus.
In the example of Fig. 17, the torus is continuously connected to 4 Dirac cones, with 4 Dirac points to close the pore, and the inner light cone is continuously connected to a Poincar\'e sphere with the smaller radius ($S_0$).
The torus ($g=1$) and the sphere ($g=0$) are describing different bulk states, respectively, while the Dirac points are edge states.
Even if we rotate the whole structure in the 3D $(S_1,S_2,S_3)$ space, the topology of these states will not be changed at all, and the polarisation independent loss merely change the scale (radius by $S_0$), such that the topology will not be changed, either.
Consequently, these topological features in the Stokes space will be robust during the propagation in the SMF, regardless of the polarisation rotations and the loss.

\begin{figure}[h]
\begin{center}
\includegraphics[width=8cm]{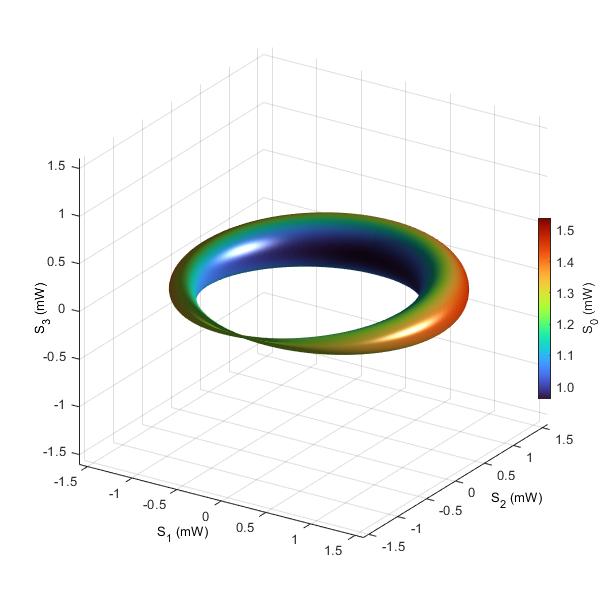}
\caption{
Closed polarisation torus.
The phase-shift is introduced upon the toroidal rotation, which induces the destructive interference, leading to the Dirac point at the anti-diagonally polarised state.
}
\end{center}
\end{figure}

Another method to close the torus is to introduce a phase-shift.
For example, we can introduce a phase-change upon the toroidal rotation as
\begin{eqnarray}
|{\rm Port \ 4^{\prime}} \rangle 
&=&
\hat{\mathcal{D}} ({\bf \hat{n}_2},\delta \phi_{\rm t}) 
\hat{\mathcal{D}} ({\bf \hat{n}_1},\delta \phi_{\rm p}) 
|{\rm Port \ 4} \rangle 
,
\end{eqnarray}
which induces the destructive rotation upon the toroidal rotation, leading to a generation of a Dirac point at the diagonally polarised state (Fig. 17).
The structure of Fig. 17 is not a torus any more, since the pore is closed, but we can introduce the toroidal phase change upon the dynamic operation \cite{Saito22h}.
Then, the polarisation torus can be dynamically switched to the conventional Poincar\'e sphere, continuously, along with the time evolution.
By considering time as an another coordinate, inspired by the time crystal \cite{Shapere12,Wilczek12}, we can dynamically switch polarisation structures with different genus.
It is important to recognise the edge must exist between different topological structures, and in order to close the torus continuously for the sphere, we need at least one Dirac point.

\subsection{Chern and Gauss-Bonnet theorems}
We consider what is the topological invariance discussed in this paper.
In physics of topological materials, the Chern number \cite{Chern46} is the topological invariance to characterise the non-trivial quantum states.
The Chern's theorem \cite{Chern46} was established through a generalisation to $\mathbb{C}$ numbers for wavefunctions \cite{Pancharatnam56,Berry84,Tomita86,Cisowski22,Moore10,Hasan10,Qi11,Nakahara90,Ando74,Ando82,Laughlin81,Kohmoto85,Hatsugai93,Murakami04,Kane05,Bernevig06,Haldane08,Armitage18}, while the classical Gauss-Bonnet theorem is valid in $\mathbb{R}$ numbers.
We focus on the polarisation torus to identify the topological invariance.

First, we evaluated the Chern number for the polarisation torus.
To calculate the Chern number, we need to integrate the Pancharatnam-Berry phase of $\gamma$ along a closed loop.
We consider the poloidal rotation, as shown in Fig. 2 (a), whose output state after the polarisation interferometer is given by
\begin{eqnarray}
|\phi_{\rm p} \rangle 
&=&
|{\rm Port \ 1}^{\prime} \rangle 
,
\end{eqnarray}
for the total rotation of $\phi_{\rm p}$, and we consider the small deviation of $\delta \phi_{\rm p}$ from $\phi_{\rm p}$, and the Berry connection \cite{Pancharatnam56,Berry84,Tomita86,Cisowski22,Nakahara90,Laughlin81,Kohmoto85,Hatsugai93}, is defined by
\begin{eqnarray}
dA
&=&
\frac{\langle \phi_{\rm p} |\phi_{\rm p} + \delta \phi_{\rm p} \rangle   -     \langle \phi_{\rm p} |\phi_{\rm p}  \rangle}
{\langle \phi_{\rm p} |\phi_{\rm p}  \rangle}
\equiv
\left \langle
\frac{\delta }
{\delta \phi_{\rm p} }
\right \rangle
\cdot
\delta \phi_{\rm p}
,
\end{eqnarray}
which yields the Pancharatnam-Berry phase as
\begin{eqnarray}
i \gamma
&=&
\oint dA
\end{eqnarray}
to give the Chern number 
\begin{eqnarray}
C=
\frac{\gamma}{2\pi}
&=&
\oint \frac{dA}{2\pi i},
\end{eqnarray}
which is the winding number for the wavefunction along the closed trajectory in the Hilbert space.
We confirmed that the Chern number for the topological torus is zero upon numerical calculations.
This is confirmed on the normalised Poincar\'e sphere of Fig. 4 (and the blue line in the inset of Fig. 12), because the circular rotation in the Stokes space along the radical direction (Fig. 2 (a)) simply corresponds to the line integration in the normalised Poincar\'e sphere, whose solid angle vanishes.
Due to the uncertainty of $4\pi$ in solid angle and the nature of the 2-level systems, the Chern number of polarisation torus is given by integers ($\mathbb{Z}$), thus, we obtain $C\in \mathbb{Z}$.
The line integration over the poloidal direction can be converted into the surface integration over the torus, such that integer Chern number characterises the nature of polarisation torus.
This is exactly the same as that for the rotationally symmetric Poincar\'e sphere, such that we have no difference in the wavefunction, which is not surprising in the definition of the Berry connection, which is defined as the overlap of the normalised wavefunctions upon a trajectory over the phase space.
If we take the integration contour over the toroidal direction, rather than the poloidal direction, the Pancharatnam-Berry phase becomes finite in agreement with the solid angle, defined by the toroidal trajectory.
However, in this case, the toroidal loop cannot cover the whole surface of the torus, such that we cannot apply the Stokes theorem to characterise the topology of the torus.
Consequently, it is reasonable to use the contour over the poloidal direction, and the Chern number of the torus is the same as that of the full sphere of the Poincar\'e sphere.
Therefore, the normalised $SU(2)$ wavefunction is not useful to characterise the polarisation torus, since the non-trivial nature of the polarisation torus in topology could be considered only when we take the variable radius of the $U(2)$ wavefunction into account for the coherent many-body states with Bose-Einstein statistics.

In fact, the topological nature of the polarisation torus is appeared in the Stokes space, where the spin expectation values could take potentially any values in $\mathbb{R}^{3}$.
For characterising the topology in the real space, we can use the Gauss-Bonnet theorem to obtain the Euler number, 
\begin{eqnarray}
\chi=2(1-g)
=
\int \frac{dS}{2\pi} K,
\end{eqnarray}
where $K$ is the product of the minimum and maximum curvatures on the surface and $dS$ is the infinitesimal surface area.
For the torus, the curvatures upon the toroidal direction changes its sign, such that the integration becomes zero, yielding $\chi=0$ and $g=1$.
The values are completely different for a sphere to have $\chi=2$ and $g=0$.
Thus, the Euler number and the genus should be appropriate as topological invariants in the polarisation torus.
Other topological features are also considered in $\mathbb{R}^{3}$, such that these numbers will be useful to consider polarisation states in the Stokes space.

\section{Conclusions}
We have shown that photons in the coherent states are described by the $U(2)$ wavefunctions, and the spin expectation values, calculated in the wavefunction, span the three dimensional Euclidean space, named the Stokes space, allowing to realise various non-trivial topological structures rather than the simple Poincar\'e sphere.
We have proposed the polarisation interferometer to realise the polarisation torus, and experimentally demonstrated the structure through the polarimetry. 
We have also shown that other topologically non-trivial structures, such as M\"obius strip, Hopf-links, and topological Dirac bosons. 
These topological structures in the Stokes space are characterised by the Euler number and genus rather than the Chern number, since the spin expectation values are observable and the proposed topological structures are realised in real values rather than complex values of wavefunctions.
We found that a bluk-edge correspondence is applicable to these topological features, and the torus and the sphere must be continuously connected, only when the Dirac point is realised at the edge to connect these structures in the Stokes space.
Topological polarisation states are robust against rotations, phase-shifts, and polarisation independent losses during the propagation in the single mode fibre, such that these features can be transmitted without breaking topological correlations.
Proposed topological structures are supported by bosonic nature of photons, allowing many photons to occupy the same state, which has remarkable difference in the fermionic Bloch state.
The energy spectrum of proposed Dirac bosons are characterised by these coherent bosons, rather than the single particle spectrum, and the linear dispersion of the energy in the bit will be observed against the helical polarisation.
We think these topological polarisation states are generic features for coherent photons emitted from ubiquitous laser sources, such that we can consider various applications such as robust optical communications and fibre sensors against signal disturbances in harsh environments or future topological quantum computing using photons.

\section*{Acknowledgements}
This work is supported by JSPS KAKENHI Grant Number JP 18K19958.
The author would like to express sincere thanks to Prof I. Tomita for continuous discussions and encouragements.

%\clearpage
\bibliography{Torus}% Produces the bibliography via BibTeX.

\end{document}